\documentclass[acmsmall, nonacm,screen,review=false]{acmart}

\usepackage [ruled, linesnumbered]{algorithm2e}
\usepackage{url}
\usepackage{mathtools}
\usepackage{multirow}
\usepackage{makecell}
\usepackage[encapsulated]{CJK}

\newtheorem{definition}{Definition}

\definecolor{textred}{RGB}{231, 29, 52}
\definecolor{textblue}{RGB}{1, 146, 213}

\setcopyright{acmlicensed}
\copyrightyear{2024}
\acmYear{2024}
\acmDOI{10.1145/3675396}

\acmJournal{TOSEM}
\acmVolume{0}
\acmNumber{0}
\acmArticle{0}
\acmMonth{7}

\title{Word Closure-Based Metamorphic Testing for Machine Translation}


\begin{document}

\author{Xiaoyuan Xie}
\authornote{Xiaoyuan Xie and Songqiang Chen are the co-corresponding authors.}
\email{xxie@whu.edu.cn}

\author{Shuo Jin}
\email{imjinshuo@whu.edu.cn}
\affiliation{\institution{School of Computer Science, Wuhan University}\country{China}}

\author{Songqiang Chen}
\authornotemark[1]
\email{i9s.chen@connect.ust.hk}

\author{Shing-Chi Cheung}
\email{scc@cse.ust.hk}
\affiliation{\institution{{Department of Computer Science and Engineering, The Hong Kong University of Science and Technology}\country{China}}}

\begin{abstract}
With the wide application of machine translation, the testing of Machine Translation Systems (MTSs) has attracted much attention. Recent works apply Metamorphic Testing (MT) to address the oracle problem in MTS testing. Existing MT methods for MTS generally follow the workflow of input transformation and output relation comparison, which generates a follow-up input sentence by mutating the source input and compares the source and follow-up output translations to detect translation errors, respectively. 
These methods use various input transformations to generate the test case pairs and have successfully triggered numerous translation errors. 
However, they have limitations in performing fine-grained and rigorous output relation comparison and thus may report many false alarms and miss many true errors. 
In this paper, we propose a word closure-based output comparison method to address the limitations of the existing MTS MT methods. 
We first propose word closure as a new comparison unit, where each closure includes a group of correlated input and output words in the test case pair. 
Word closures suggest the linkages between the appropriate fragment in the source output translation and its counterpart in the follow-up output for comparison.
Next, we compare the semantics on the level of word closure to identify the translation errors. 
In this way, we perform a fine-grained and rigorous semantic comparison for the outputs and thus realize more effective violation identification.
We evaluate our method with the test cases generated by five existing input transformations and the translation outputs from three popular MTSs. 
Results show that our method significantly outperforms the existing works in violation identification by improving the precision and recall and achieving an average increase of 29.9\% in F1 score. It also helps to increase the F1 score of translation error localization by 35.9\%.
\end{abstract}


\maketitle

\begin{CJK*}{UTF8}{gbsn}
\section{Introduction}

Machine translation systems (MTSs) have seen more and more applications in our daily life.
Many popular MTSs, such as Google Translate and Bing Microsoft Translator, are providing translation services for millions of users every day~\cite{GoogleTrans-user,BingTrans-user}. However, the automatically generated translations are not always correct. If not recognized, the incorrect translations provided by MTS can lead to significant issues, such as political conflicts~\cite{mistranslation-political}, false arrest~\cite{mistranslation-arrest1,mistranslation-arrest2}, financial loss~\cite{mistranslation-financial1,mistranslation-financial2}, and vaccine mishap~\cite{mistranslation-vaccine}.
In order to avoid these problems, there have been a few testing methods proposed for MTS to detect its potential mistranslations. 

One promising idea to efficiently test MTS is to apply Metamorphic Testing (MT) \cite{MT}. Some early methods adopt the round-trip translation to construct follow-up cases \cite{RTT2005, RTT2006, RTT2007, semmt}.
Recently, a few new methods propose more sophisticated Metamorphic Relations (MRs) that manipulate particular components of an input sentence to generate follow-up cases \cite{sit,purity,patinv,cat,cit}.
In the output comparison, these new methods check whether there are cases where the same (or different) input word is translated into different (or similar) meanings. If yes, there is a violation being reported, indicating at least one translation error. 
These new methods include SIT \cite{sit}, CAT \cite{cat}, Purity \cite{purity}, CIT \cite{cit} and PatInv \cite{patinv}. Test cases generated by the input transformations in these five methods successfully trigger numerous mistranslations of MTS.

However, these methods still have several limitations regarding ``\textit{what to compare}'' and ``\textit{how to compare}''. In terms of \textit{``what to compare''}, we identified three major limitations: 
(1) Some methods cannot compare two output translations at a fine granularity for precise comparison. Specifically, SIT and CAT detect the violation according to the similarity between the entire (estimated) unchanged part of the two translations. However, such an overall similarity can not effectively reflect the fine-grained differences between the translations. As a result, violations can be incorrectly revealed on some correct translations while some translation errors can be missed. 
For instance, our experiments conducted on the English-to-Chinese translation service of Google Translate show that 48.0\% and 40.6\% of false alarms and 41.2\% and 100.0\% of missing detections in SIT and CAT can be attributed to this limitation.
(2) Some methods lack the linkage between the fragment in the source output and its counterpart in the follow-up output to implement rigorous comparison. Specifically, without such linkages, Purity and CIT have to adopt relatively loose comparison strategies to judge the correctness of outputs, which may miss many translation errors. 
For example, we have found that 42.9\% and 31.7\% of missing detections in English-to-Chinese translations from Google Translate can be due to this limitation for Purity and CIT, respectively.
(3) Some method (i.e., PatInv) ignores to detect the incorrect translations of the unchanged input words, which may also miss many translation errors. 
Take the testing of the English-to-Chinese translation from Google Translate as an example, all the violations missed by PatInv can be attributed to this limitation according to our experiment.
In terms of \textit{``how to compare''}, we summarized two limitations: 
(1) Some methods that apply structure-based comparison methods fail to detect the incorrect translation that has the same structure but incorrect semantic meanings as its counterpart. 
Based on our experiment, when testing Google Translate to translate English into Chinese, this limitation causes 52.9\% and 68.3\% of missing detections for SIT and CIT, respectively.
(2) Some methods adopt text-based comparison methods and thus cannot recognize the synonyms between the two translations. 
For example, in our experiment on the English-to-Chinese translations of Google Translate, we found that 72.5\% and 88.2\% of false alarms for CAT and Purity, respectively, can be attributed to this limitation.
Thus, these limitations are pervasive and largely hinder the effectiveness of current methods.

Meanwhile, it is challenging to address the aforementioned limitations and achieve rigorous output violation checking. Specifically, it is necessary to link the fine-grained counterparts in the source and follow-up translations, which will be the proper comparison subjects against the metamorphic relation.
To achieve a fine-grained violation identification, the native grammar unit, such as clause, phrase or word, is not suitable to work as the counterparts since they are not flexible and hard to match.
Besides, to achieve a rigorous comparison, building linkages between these counterparts is challenging due to the poor performance of word-level alignment tools.

In this work, we propose to perform MT for MTS based on a new concept called \textbf{\textit{word closure}} to alleviate the limitations in the existing output comparison methods. 
We first rule out the coarse comparison on the overall similarity of the outputs. The core idea of our solution is to \textbf{link} each appropriate fragment in the source output and its counterpart in the follow-up output (i.e. to construct word closures), such that a finer-grained and rigorous comparison can be realized. In this way, we can properly address the limitations of ``what to compare''.
Regarding the limitations of ``how to compare'', we compare the translations by evaluating their semantics to address the limitations of the structure- and text-based methods.
With the help of word closure and semantic comparison, we can detect many missed errors (i.e. false negatives) and incorrectly detected errors (i.e. false positives) in existing works. In addition, the merit of ``fine granularity'' enables our method to locate the fine-grained violations more precisely, i.e., the specific words that cause the overall violation within the test case pair. 

We perform a comprehensive evaluation of our proposed output comparison approach to detect translation errors from three popular MTSs. 
The experiment results show that our approach can effectively reduce the false negatives and false positives in the existing methods, and lead to a significant improvement in the overall performance. Specifically, the average increase in F1 score is as high as 29.9\% as compared with the five existing methods.
For fine-grained violation locating, our approach outperforms the existing methods with an average increase in F1 score of 35.9\%, which demonstrates the advantage of our method in locating the specific words that cause the overall violation within the test case pair.

To summarize, this paper addresses the limitations of violation identification methods applied by the existing MT testing works for MTS to achieve better performance in translation error detection. Our contributions are outlined below:

\begin{itemize}
\item{We review the latest MTS metamorphic testing works whose MRs mutate the source input sentence,  summarize the primary limitations in their output relation comparison methods, and show the impact of such limitations on the error detection performance.}

\item {We propose a word closure-based output comparison method to alleviate the limitations of existing works in violation identification.
Our approach can link each appropriate fragment in the source output and its counterpart in the follow-up output (i.e. to construct word closures), such that a finer-grained and rigorous comparison can be realized. 
Combined with semantic comparison, our method can properly resolve the problems of ``what to compare'' and ``how to compare''.}

\item{The built linkages between the fragments in the source translation and their counterparts in the follow-up translation via word closure can further support the fine-grained violation localization, which can pinpoint the specific words giving rise to the detected violations.}

\item{The experiment results show that our word closure-based comparison method is very effective in reducing the false positives and false negatives in existing methods in both the widely-studied English-Chinese translation and the unexplored essential Chinese-English translation scenarios, and hence leads to a significant improvement in the overall violation detection ability. Specifically, our method achieves an average increase of 29.9\% in the F1 score as compared with existing methods. 
In terms of locating fine-grained violations, our approach achieves an average increase of 35.9\% in F1 score than the existing methods.}
\end{itemize}

\textbf{The dataset, replication package, and supplementary materials for this paper are available online at \cite{DATARELEASE}.}

The remaining paper is organized as follows. In Section~\ref{sec:background}, we review the five target latest MT methods for MTS testing. 
Section~\ref{sec:motivation} summarizes the limitations of existing work and illustrates specific examples, which motivate our work.
In Section~\ref{sec:methodology}, we introduce our word closure-based output comparison approach and explain its detailed implementations. 
Section~\ref{sec:expsetup} lists the experimental setup and Section~\ref{sec:expresult} presents the experiment results and analysis for the research questions. 
In Section~\ref{sec:discussion}, we study the helpfulness of our approach in enhancing the robustness of the machine translation model, summarize the limitations of our approach and outline some potential directions for future advancements.
Section~\ref{sec:threatstovalidity} discusses the threats to validity of this work, and Section~\ref{sec:relatedwork} overviews the related works.
Finally, we conclude the paper in Section~\ref{sec:conclusion}.

\section{Background}
\label{sec:background}

Recent automated MTS testing methods~\cite{RTT2005, RTT2006, RTT2007, semmt,MT4MT,sit,purity,patinv,transrepair,cat,cit} are mostly based on Metamorphic Testing (MT).
MT solves the oracle problem of MTS testing via Metamorphic Relations (MRs), which describe the necessary properties of MTS in relation to multiple inputs and their expected outputs \cite{MT_Review}. The violation of the MRs indicates the incorrect translation of MTS.
Some early proposed methods test MTS via round-trip translation \cite{RTT2005, RTT2006, RTT2007, semmt}. Recently, some new methods focus on MRs that manipulate particular components of an input sentence \cite{MT4MT,sit,purity,patinv,transrepair,cat,cit}. 
Specifically, these MRs consist of two primary steps to test MTS, i.e., the input transformation (denoted as \textbf{IT}) and the output relation comparison (denoted as \textbf{OR}). Given a source input sentence (denoted as $S_s$), the testing method transforms $S_s$ into a new follow-up input sentence (denoted as $S_f$) according to the input transformation of the MR. Next, the translations of both $S_s$ and $S_f$, denoted as $T_s$ and $T_f$, respectively, are derived by running MTS with $S_s$ and $S_f$. 
Finally, $T_s$ and $T_f$ are compared to check if $T_s$ and $T_f$ 
violate the expected output relation defined by the MR. If a violation is identified, there should be translation errors within $T_s$ and $T_f$. In this paper, we propose an approach to effectively enhance the MTS testing methods that transform the input sentence, which are more popular recently. 

In the family of MT-based MTS testing methods \textbf{that transforms the input sentence}, five recent methods, i.e., \textbf{SIT}~\cite{sit}, \textbf{PatInv}~\cite{patinv}, \textbf{Purity}~\cite{purity}, \textbf{CIT}~\cite{cit}, and \textbf{CAT}~\cite{cat}, have demonstrated outstanding performance in detecting incorrect translations of MTS. 
In the following, we introduce the MRs of these methods: 

\begin{enumerate}
\item{\textit{\textbf{SIT:}} SIT is based on the MR that the translations of similar sentences should exhibit similar sentence structure. It replaces one word in $S_s$ with another word that keeps the same part of speech to generate a similar follow-up sentence $S_f$. We denoted this input transformation as \textbf{\mbox{IT-1}}. Next, SIT compares whether the constituency parse tree or dependency parse tree of two output translations $T_s$ and $T_f$ are similar. This output relation is denoted as \textbf{OR-1}. If the distance between the two parse trees exceeds the predefined threshold, it reports this pair of test cases as a violation.}

\item{\textit{\textbf{CAT:}} 
CAT employs the MR that the translations of two sentences with one different word should be similar except for those of the different words. CAT replaces one word in $S_s$ with another word similar in semantics to generate the $S_f$ (\textbf{IT-2}). To check whether the two input sentences are translated similarly except for the replaced words (\textbf{OR-2}), CAT leverages a strategy to roughly remove the translations of the replaced words from comparison and calculates the similarity score between the estimated unchanged part of the two translations. If the overall similarity score is below the predefined threshold, CAT reports a violation as the translations of the same input words are largely different.} 

\item{\textit{\textbf{Purity:}}\label{back-patinv} Purity is based on the MR that a noun phrase should be translated similarly in different contexts. Purity proposes the input transformation to extract a noun phrase from $S_s$ as the follow-up input $S_f$ (\textbf{IT-3}). 
To check whether the extracted noun phrase is translated similarly in $T_s$ and $T_f$ (\textbf{OR-3}),  Purity measures how many word occurrences are in $T_f$ but not in $T_s$ as the distance between the source and follow-up translations of the identical noun phrase. If the distance is above the threshold, Purity considers that the target noun phrase is translated differently in $T_f$ and $T_s$ and reports this test case pair as a violation.}

\item{\textit{\textbf{CIT:}} The MR of CIT is that the constituency structure of a sentence's translation should remain unchanged if an adjunct, optional constituent of a sentence, is inserted in the sentence. The input transformation proposed by CIT is to add an adjunct into $S_s$ to generate the $S_f$ (\textbf{IT-4}). Given an original sentence, CIT leverages the sentence compression tool, which can automatically compress a sentence by removing all the adjuncts from it, to identify the adjuncts in the sentence. With the compressed sentence as $S_s$, CIT adds one identified adjunct to generate the $S_f$. To verify whether the constituency structure of $T_s$ remains unchanged in $T_f$ (\textbf{OR-4}), CIT checks whether all the paths in the constituency parse tree of $T_s$ exist in that of $T_f$. CIT will report a violation if any path violates the above expected output relation.}

\item{\textit{\textbf{PatInv:}} PatInv generates sentences with different meanings to detect translation errors. The MR of PatInv is that sentences with different meanings should have different translations. PatInv proposes two implementations to change the meaning of the source input sentence $S_s$ to generate the $S_f$ (\textbf{IT-5}). The first method is to replace one word in $S_s$ with a word of different semantics. Another one is to remove a meaningful word or phrase from $S_s$. As $S_s$ and $S_f$ have different meanings, their translations $T_s$ and $T_f$ are supposed to be different (\textbf{OR-5}). If $T_s$ and $T_f$ are the same, PatInv will report a violation.}

\end{enumerate}

\begin{table*}
\addtolength{\tabcolsep}{-1pt}
\centering\scriptsize
    \caption{Taxonomy of the five metamorphic testing methods for MTS.}
    \label{taxonomy}
    \begin{tabular}{c|c|l|l|l}
        \hline
        \textbf{\makecell[c]{Output \\ Comparison}} & \textbf{\makecell[c]{Input \\ Transformation}} & \multicolumn{2}{c|}{\textbf{Approach}} & \textbf{\makecell[c]{Summary of Output Comparison}} \\
        \hline
        \multirow{21}{*}{\makecell[c]{Equivalence \\ comparison}} & \multirow{12}{*}{\makecell[c]{Word \\ replacement}} & \multirow{12}{*}{\rotatebox{90}{Class-A}} & \multicolumn{1}{c|}{\textbf{SIT}} & \multirow{12}{*}{\makecell[l]{Compare \textbf{the overall similarity} \\ between the entire (estimated) \\ unchanged part of the two transla- \\ tions to detect the violation.}} \\
        &&&\multirow{4}{*}{\makecell[l]{\textbf{IT-1}: Replace one word in $S_s$ with another word of \\ the same part of speech to generate $S_f$. \\ \textbf{OR-1}: The constituency parse trees or dependency \\ parse trees of $T_s$ and $T_f$ should be the same.}}&\\
        &&&&\\
        &&&&\\
        &&&&\\
        &&&\multicolumn{1}{c|}{\textcolor{textred}{\textit{structure-based comparison}}}&\\
        \cline{4-4}
         & & & \multicolumn{1}{c|}{\textbf{CAT}} & \\
        &&&\multirow{4}{*}{\makecell[l]{\textbf{IT-2}: Replace one word in $S_s$ with another word \\ of similar context-aware semantics to generate $S_f$. \\ \textbf{OR-2}: The same input words should be translated \\ similarly in $T_s$ and $T_f$.}}&\\
        &&&&\\
        &&&&\\
        &&&&\\
        &&&\multicolumn{1}{c|}{\textcolor{textblue}{\textit{text-based comparison}}}&\\
        \cline{2-5}
         & \multirow{9}{*}{\makecell[c]{Decremental \\ or Incremental \\ transformation}} & \multirow{9}{*}{\rotatebox{90}{Class-B}} & \multicolumn{1}{c|}{\textbf{Purity}} & \multirow{9}{*}{\makecell[l]{Check the inclusive relation between \\ the two translations \textbf{token by token}.}} \\
        &&&\multirow{2}{*}{\makecell[l]{\textbf{IT-3}: Extract a noun phrase from $S_s$ as $S_f$.  \\ \textbf{OR-3}: The words in $T_f$ should exist in $T_s$.}}&\\
        &&&&\\
        &&&\multicolumn{1}{c|}{\textcolor{textblue}{\textit{text-based comparison}}}&\\
        \cline{4-4}
        & & & \multicolumn{1}{c|}{\textbf{CIT}} & \\
        &&&\multirow{3}{*}{\makecell[l]{\textbf{IT-4}: Insert an adjunct into $S_s$ to generate $S_f$. \\ \textbf{OR-4}: Every path in the constituency parse tree of \\ $T_s$ should exist in that of $T_f$.}}&\\
        &&&&\\
        &&&&\\
        &&&\multicolumn{1}{c|}{\textcolor{textred}{\textit{structure-based comparison}}}&\\
        \hline
        \multirow{6}{*}{\makecell[c]{Unequivalence \\ comparison}} & \multirow{6}{*}{Semantic change} & \multirow{6}{*}{\rotatebox{90}{Class-C}} & \multicolumn{1}{c|}{\textbf{PatInv}} & \multirow{6}{*}{\makecell[l]{Only check if the \textbf{changed input} \\ \textbf{words} are translated differently.}} \\
        &&&\multirow{3}{*}{\makecell[l]{\textbf{IT-5}: Replace one word in $S_s$ with another word \\ of different meanings, or removing a meaningful \\ word or phrase from $S_s$ to generate $S_f$. \\ \textbf{OR-5}: $T_s$ and $T_f$ should be different.}}&\\
        &&&&\\
        &&&&\\
        &&&&\\
        &&&\multicolumn{1}{c|}{\textcolor{textblue}{\textit{text-based comparison}}}&\\
        \hline
    \end{tabular}
\end{table*}

As shown in Table~\ref{taxonomy}, the five testing methods for MTS can be divided into categories according to their output comparison strategies and input transformations.
\textbf{Based on the output comparison strategy}, the five testing methods are divided into two general categories, i.e., ``Equivalence comparison'' (\textbf{Class-A and Class-B}) including SIT, CAT, Purity, and CIT, and `Unequivalence comparison'' (\textbf{Class-C}) including PatInv. 
Within ``Equivalence comparison'', the four methods can be further divided into \textbf{two sub-categories according to their input transformations.} SIT and CAT are included in ``Word replacement'' (\textbf{Class-A}). They both replace one word in the source input sentence to generate a similar follow-up input sentence. 
Although Purity and CIT apply two seemingly different input transformations, i.e., decremental and incremental transformations, they both build a pair of input sentences with different lengths, wherein the shorter sentence is a part of the longer one. Thus, we categorize them together as \textbf{Class-B}.
For clarity, we summarize the output relation comparison method of these three categories in the last column of Table~\ref{taxonomy}.

Besides, the five testing methods can also be categorized based on their strategies in evaluating the similarity (or dissimilarity) between the output translations. 
Specifically, SIT and CIT use ``\textbf{structure-based comparison}'' and CAT, Purity, and PatInv use ``\textbf{text-based comparison}''.

However, all these categories of MTS testing methods have limitations in their output comparison methods. Such limitations will be discussed in Section~\ref{sec:motivation}.

\section{Motivation}
\label{sec:motivation}

The five metamorphic testing methods for MTS have successfully detected mistranslations of MTS with their input transformations and output relations. 
However, \textbf{the identified mistranslations based on MR violation can be wrong} due to \textbf{the limitations of their output relation comparison methods}. 
The wrong identification results can be categorized into \textit{False Negatives} (FNs) and \textit{False Positives} (FPs). 
An FP means a pair of test cases, whose outputs satisfy the expected output relation of MR, is misidentified as a violation. 
Conversely, if a true violation is missed, an FN will be counted. In the following, we illustrate each limitation with practical examples
\footnote{For more detailed explanations of the examples (e.g., Chinese meanings), please refer to our online artifact~\cite{DATARELEASE}.}.

\begin{figure*}[!htb]
\centering
\includegraphics[width=0.88\linewidth]{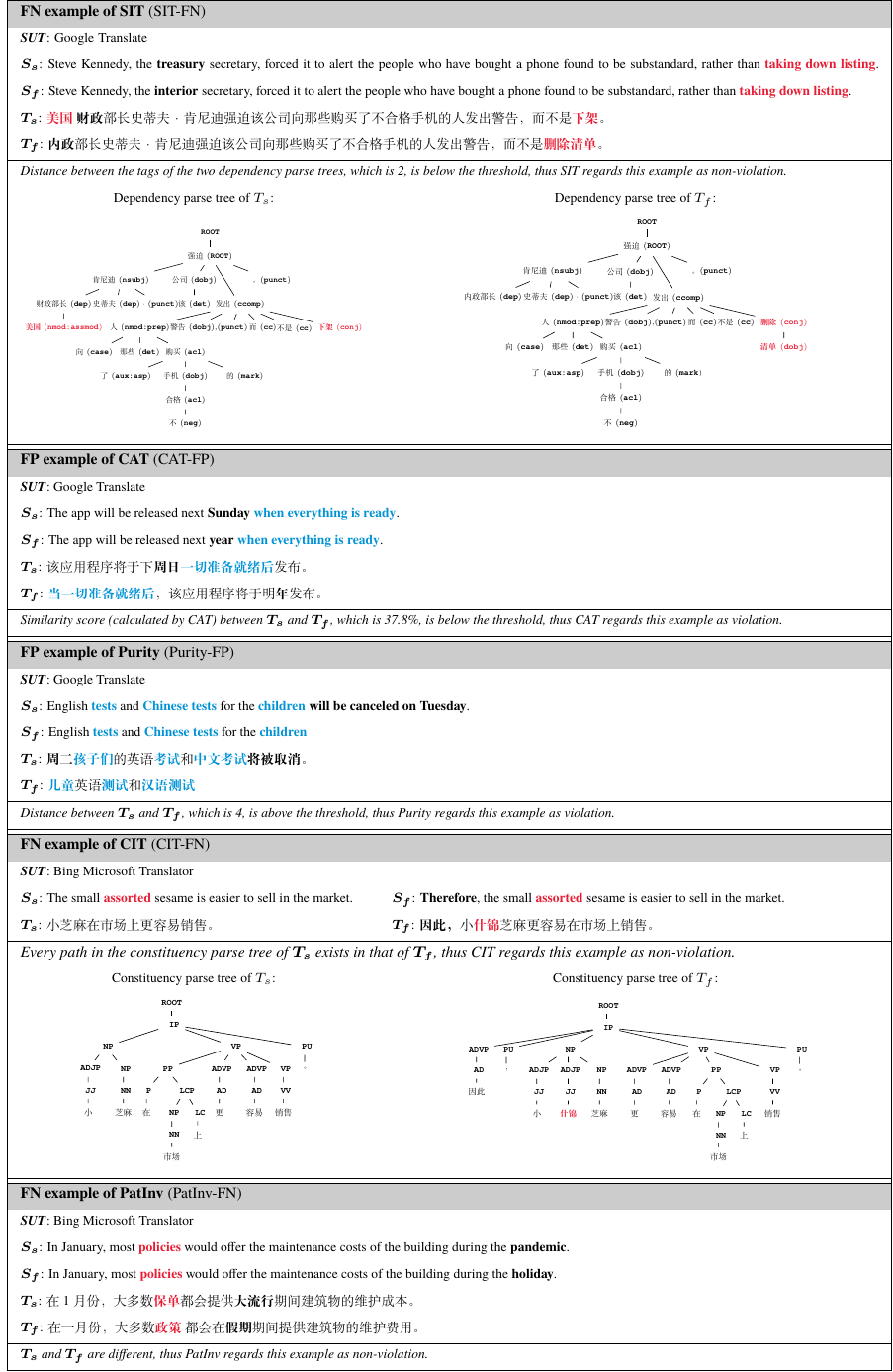}
\caption{Examples of the limitations of the existing output comparison methods.}
\label{examples}
\end{figure*}

\begin{enumerate}

\item{\textbf{Comparison methods of Class-A cannot compare two output translations at a fine granularity for precise comparison.} 
SIT and CAT compare the overall sentence similarity between the entire (estimated) unchanged part of $T_s$ and $T_f$ to detect the violation. Specifically, SIT checks whether the two sentences are translated similarly by computing the distance between the entire $T_s$ and the entire $T_f$. Meanwhile, although CAT approximately removes the translations of the mutated words from the comparison, it computes the overall similarity score between the estimated unchanged part of translations in $T_s$ and $T_f$ to detect the violation.
If the overall distance or similarity cannot satisfy the preset threshold, they regard this pair of test cases as a violation. 
However, such a coarse comparison strategy will inevitably introduce some FPs or FNs.
On one hand, they can miss some finer violations, which do not lead to an obvious decrease in the overall similarity between $T_s$ and $T_f$. This will lead to FNs. 
An FN example of SIT is shown in Fig.~\ref{examples}. SIT generates $S_f$ by replacing the word ``treasure'' in $S_s$ with ``interior'', as marked in bold.  In the translations $T_s$ and $T_f$, there is a redundant translation ``美国'' (the U.S.) in $T_s$, and the same input words ``taking down listing'' in $S_s$ and $S_f$ are translated as different meanings in $T_s$ and $T_f$, i.e., ``下架'' (removed from shelves) and ``删除清单'' (delete list). Therefore, this example is indeed a violation. 
However, due to the small difference introduced by the above translation error, the overall distance score between the parse trees of $T_s$ and $T_f$ is lower than the threshold defined by SIT.
As a result, the output translations for this test case pair are identified as non-violation by SIT. Similar FN cases can be found in CAT as well. 
On the other hand, coarse-grained comparison may misidentify the test cases that satisfy the output relation as a violation, which leads to FPs. 
An FP example of CAT is shown in Fig.~\ref{examples}. CAT generates $S_f$ by replacing the word ``Monday'' in $S_s$ with ``Sunday'', as marked in bold. Although the translation of the adverbial clause ``when everything is ready'' is placed at different positions in $T_s$ and $T_f$, as highlighted in blue, there is no violation within $T_s$ and $T_f$ as they share similar meanings~\cite{adverb-position}. However, CAT misidentifies it as a violation due to the overall low similarity between $T_s$ and $T_f$ caused by the different positions of the adverbial clause. Similar FPs can also be found in SIT.
}

\item{\textbf{Comparison methods of Class-B lack the linkage between the fragment in the source output and its counterpart in the follow-up output to implement rigorous comparison.} 
Given a sentence to translate, Purity and CIT generate a new input sentence by adding or deleting contents. Unlike Class-A methods that use the overall similarity to detect the translation errors for two relatively similar input sentences, Purity and CIT judge the inclusion relationship between the outputs for a pair of inputs.
Specifically, they check whether the words or syntax parse tree paths of the shorter translation exist in that of the longer one token by token.
However, this inclusion relationship checking method is relatively loose.
It can lead to FNs in some cases when the shorter translation misses some necessary content or the longer translation contains redundant content.
An FN example of CIT is shown in Fig.~\ref{examples}. CIT generates $S_f$ by inserting the adjunct ``Therefore,'' into $S_s$, as marked in bold. Because the same input word ``assorted'' is translated as ``什锦'' in $T_f$ but missed in $T_s$, there is a violation over this test case pair. 
However, CIT identifies this test case pair as non-violation because the constituency parse tree of $T_f$ includes every path in the constituency parse tree of $T_s$. As a result, lacking the linkages between the counterparts between $T_s$ and $T_f$, CIT implements a relatively loose comparison strategy and fails to detect the violation in this example.
Similar FNs can be found in Purity as well.
}

\item{\textbf{Comparison method of Class-C ignores to detect the incorrect translations of the unchanged input words.}
PatInv only compares whether the altered words in $S_s$ and $S_f$ are translated differently while ignoring the correctness of the other translations. As a result, PatInv fails to detect the incorrect translations in $T_s$ and $T_f$ of the same input words, which leads to FNs. An FN example of PatInv is shown in Fig.~\ref{examples}. PatInv generates $S_f$ by replacing the word ``pandemic'' in $S_s$ with ``holiday'', as marked in bold. Although the same input word ``policies'' is translated as different meanings in $T_s$ and $T_f$, i.e., ``保单'' (guarantee slip) and ``政策'' (policies), according to its output checking rule, PatInv regards this example as non-violation just because $T_s$ and $T_f$ are different.
}

\end{enumerate}

\textbf{As a reminder, apart from the limitations from the above ``what to compare'' aspect, current solutions to ``how to compare'' also have problems.}

First, {\textbf{the structured-based comparison methods fail to detect the incorrect translation that has the same structure as its counterpart.} Incorrect translation within $T_s$ and $T_f$ do not always lead to the different structures between $T_s$ and $T_f$. Therefore, structured-based comparison methods can miss many incorrect translations, leading to FNs.}

Secondly, {\textbf{the text-based comparison methods cannot recognize the synonyms between the two translations.} 
Text-based comparison methods, which adopt string-based metrics (e.g. BLEU), measure the similarity between two translations based on the plain lexicographic match without considering meaning~\cite{string-based-metric}. As a result, CAT and Purity can misidentify the synonyms within $T_s$ and $T_f$ as violations, leading to FPs. For PatInv, which reports a violation only if the two translations are the same, the test case pairs with violations may be ignored due to the differences involved by synonyms, leading to FNs.
To illustrate the limitation of the text-based comparison methods, we present an FP example of Purity in Fig.~\ref{examples}. Purity extracts a noun phrase from $S_s$ as the $S_f$. Although there are many synonyms between $T_s$ and $T_f$, e.g., ``考试'' (tests) in $T_s$ and ``测试'' (tests) in $T_f$, there is no violation between $T_s$ and $T_f$ because these synonyms share similar semantic meanings. Unable to recognize these synonyms, Purity misidentifies this example as a violation because four words from $T_f$ can not be found in $T_s$, the sum of which is above the threshold.
}

\begin{table}[ht]
\centering\footnotesize
    \caption{Limitations of the existing output relation comparison methods.}
    \label{limitation}
    \begin{tabular}{c|c|l|c}
        \hline
        \multicolumn{2}{c|}{\textbf{\makecell[c]{Approach}}} & \textbf{\makecell[c]{Limitation}} & \textbf{\makecell[c]{Result}} \\
        \hline
        \multirow{4}{*}{\rotatebox{90}{Class-A}}&\multirow{2}{*}{\makecell[c]{\textbf{SIT}}} & \multirow{4}{*}{\makecell[l]{Cannot compare two output translations at a fine granularity for precise compa-\\rison.}}
        & \multirow{4}{*}{\makecell[c]{FP+FN}} \\
        &&& \\
        \cline{2-2}
        &\multirow{2}{*}{\makecell[c]{\textbf{CAT}}} &  & \\
        &&& \\
        \cline{1-4}
        \multirow{4}{*}{\rotatebox{90}{Class-B}}&\multirow{2}{*}{\makecell[c]{\textbf{Purity}}} & \multirow{4}{*}{\makecell[l]{Lack the linkage between the fragment in the source output and its counterpart \\ in the follow-up output to implement rigorous comparison.}} & \multirow{4}{*}{\makecell[c]{FN}} \\
        &&& \\
        \cline{2-2}
        &\multirow{2}{*}{\makecell[c]{\textbf{CIT}}} &  & \\
        &&& \\
        \hline
        \multirow{4}{*}{\rotatebox{90}{Class-C}}&\multirow{4}{*}{\makecell[c]{\textbf{PatInv}}} & \multirow{4}{*}{\makecell[l]{Fail to detect the incorrect translations of the same input words.}} & \multirow{4}{*}{\makecell[c]{FN}} \\
        &&& \\
        &&& \\
        &&& \\
        \hline
        \hline
        \multirow{4}{*}{\rotatebox{90}{\textcolor{textred}{\makecell[c]{Structure\\-based}}}}&\multirow{2}{*}{\makecell[c]{\textbf{SIT}}} & \multirow{4}{*}{\makecell[l]{Fail to detect the incorrect translations that have the same structure as its coun- \\ terpart.}} & \multirow{4}{*}{\makecell[c]{FN}} \\
        &&& \\
        \cline{2-2}
        &\multirow{2}{*}{\makecell[c]{\textbf{CIT}}} &  & \\
        &&& \\
        \cline{1-4}
        \multirow{3}{*}{\rotatebox{90}{\textcolor{textblue}{\makecell[c]{Text\\-based}}}}&\multirow{1}{*}{\makecell[c]{\textbf{CAT}}} & \multirow{3}{*}{\makecell[l]{Cannot recognize the synonyms between the two translations.}} & \multirow{2}{*}{\makecell[c]{FP}} \\
        \cline{2-2}
        &\textbf{Purity} &  & \\
        \cline{2-2}
        \cline{4-4}
        &\textbf{PatInv} &  & FN \\
        \hline
    \end{tabular}
\end{table}

To summarize, every existing output comparison method has at least two of the above limitations, from the aspects of both ``what to compare'' and ``how to compare''. These limitations, as summarized in Table~\ref{limitation}, may lead to FPs, FNs, or both and therefore degrade the testing method performance.
On one hand, reporting many FPs reduces the reliability of the test results and requires additional manual filtering to identify true translation errors.
On the other hand, some serious translation errors within the FNs may be undetected and unfixed, posing potential risks to the practical application of MTS. Furthermore, too many FNs may even mislead users into believing that the quality of MTS is dependable and using MTS without hesitation in some key areas.
In order to achieve more adequate violation identification, in this work, we propose a word-closure-based output comparison method to address the aforementioned limitations.

\section{Methodology}
\label{sec:methodology}

\subsection{Core Idea and Approach Overview}

In this work, we propose a new output relation comparison approach for Metamorphic Testing of Machine Translation. It addresses the limitations regarding ``what to compare'' and ``how to compare'' of the output checking method in existing methods introduced in Section~\ref{sec:motivation}. It realizes more rigorous output checking and can be used with existing input transformation methods.

First, to address \textbf{the limitations from ``what to compare'',} we propose a new type of granularity, namely ``word closure''. We perform output comparisons on this granularity level.
As mentioned in Section~\ref{sec:motivation}, it is too coarse to compare the source and follow-up outputs by measuring the overall similarity between translations. 
A finer-grained entity is needed for comparison. Meanwhile, simply adopting native grammar units such as words, phrases, or clauses cannot meet the requirement in our MT scenario. The main reason is that, in MT, test cases are changed via various transformations. As a result, we usually need to compare a particular fragment in the source output and its \textbf{\textit{counterpart}} in the follow-up output against some relations (i.e. MRs). 
In other words, the core challenge during this process is to \textbf{identify the corresponding relation between a particular fragment in the source output and its counterpart in the follow-up output}. Comparisons against MRs should be performed by referring to these relations.
Simply adopting the native grammar units cannot achieve this goal, as these units across the source and follow-up outputs do not always show explicit corresponding relations.
In this work, we propose ``\textbf{\textit{word closure}}'' to address the above challenge. A word closure comprises the appropriate counterparts in a pair of source and follow-up outputs, which are linked for comparison against the output relation, as well as their corresponding fragments in the source and follow-up inputs.
Section~\ref{sec:wordClosure} will explain the details of word closure. 

To construct the word closures, word mappings between the input sentences and output translations are necessary. It is not difficult to think of adopting a general word alignment tool, which is also mentioned in \cite{purity, transrepair}. However, the accuracy of basic word alignments does not meet our requirements. Thus, we propose two algorithms to refine the word alignment results, which are described in Section~\ref{subsec:wordalignment-refine}. 
After the construction of all word closures for a pair of metamorphic test cases, comparisons against the MR will be performed between the translation fragments linked by word closures.

\begin{figure*}[!th]
\centering
\includegraphics[width=\linewidth]{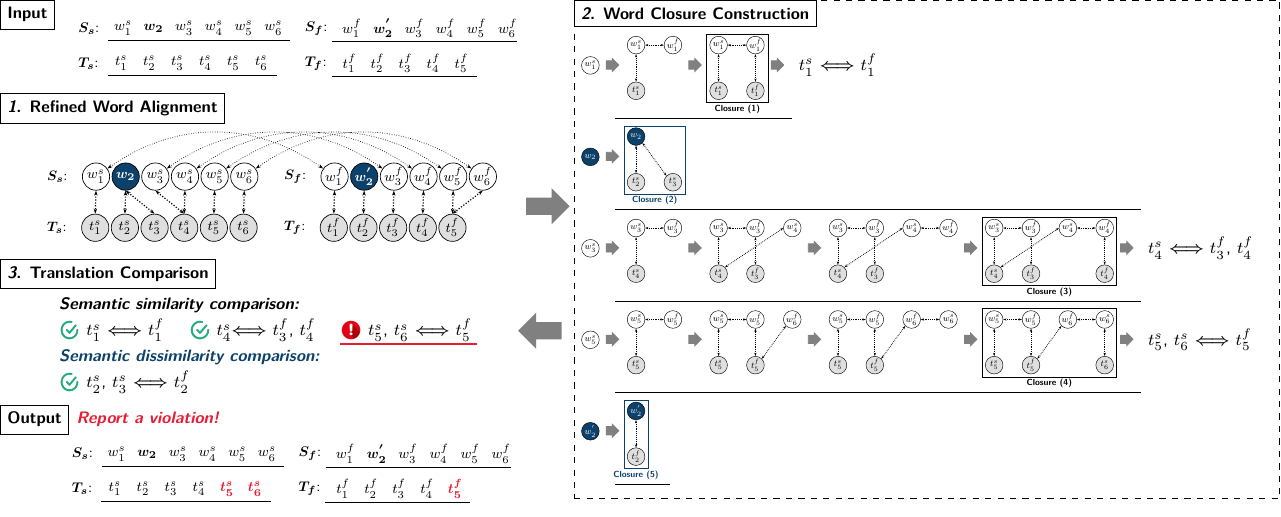}
\caption{Overview of our approach.}
\label{overview}
\end{figure*}

The idea of performing comparison within the linked counterparts in the source and follow-up outputs based on word closures allows us to address the limitations of Class-A, Class-B, and Class-C. 
\textit{Firstly}, we use word closure to split the translations into finer granularity, which is shorter than the comparison unit of entire translations adopted by Class-A. This helps to avoid the FPs and FNs introduced by a coarse comparison granularity.
\textit{Secondly}, word closures can link each fragment from $T_s$ with its counterpart in $T_f$. This enables a more rigorous comparison by clearly identifying the elements to compare, which cannot be realized by Class-B. 
In addition, Class-C only compares the translations of changed words. Meanwhile, word closures can be utilized to compare the translations of both the same input words and the changed input words to detect the incorrect translations more adequately.

By considering \textbf{the limitations from ``how to compare''}, we propose to compare the semantic similarity (or dissimilarity if needed) of the translations to detect violations. 
On one hand, comparing the semantic similarity can expose the violations with different meanings but identical structures, which are ignored by the structure-based comparison methods. On the other hand, semantic-based comparison can identify the synonyms within the source and follow-up translations due to their similar semantics, thereby avoiding the FPs and FNs caused by the text-based comparison methods. 
When comparing translations of the same input words, we detect violations based on their semantic similarity. 
For the translations of the changed input words, we check their semantic dissimilarity (similarity) if the changed input words have different (similar) meanings.
As a reminder, in our approach, we compare the semantic similarity at a fine granularity, i.e., word closure. By properly addressing the limitations about ``what to compare'', the semantic comparison can be performed more accurately.

Table~\ref{solution} briefly summarizes how our method addresses each limitation of the existing output comparison methods for MTS testing.

\begin{table}[h!]
\centering\footnotesize
    \caption{Our solutions to the limitations of existing output comparison methods.}
    \label{solution}
    \begin{tabular}{c|c|l}
        \hline
        \multicolumn{2}{c|}{\textbf{Approach}}& \multicolumn{1}{c}{\textbf{Our Solution}} \\
        \hline
        \multirow{4}{*}{\rotatebox{90}{Class-A}}& \multirow{2}{*}{\makecell[c]{\textbf{SIT}}} & \multirow{8}{*}{\makecell[l]{We propose word closure to split the source and follow-up translations into \textbf{finer and appropriate} \\ \textbf{granularity}, and build the \textbf{linkage} between the fragment in the source output with its counterpart \\  in the follow-up output for rigorous output relation comparison.}} \\
        && \\
        \cline{2-2}
        & \multirow{2}{*}{\makecell[c]{\textbf{CAT}}} & \\
        && \\
        \cline{1-2}
        \multirow{4}{*}{\rotatebox{90}{Class-B}} & \multirow{2}{*}{\makecell[c]{\textbf{Purity}}} & \\
        && \\
        \cline{2-2}
        & \multirow{2}{*}{\makecell[c]{\textbf{CIT}}} & \\
        && \\
        \hline
        \multirow{4}{*}{\rotatebox{90}{Class-C}} & \multirow{4}{*}{\makecell[c]{\textbf{PatInv}}} & \multirow{4}{*}{\makecell[l]{In addition to comparing whether \textbf{the different} \textbf{input words} are translated differently (only for \\ IT-5), we also compare whether \textbf{the same input} \textbf{words} are translated similarly to detect the \\ violation within them.}} \\
        && \\
        && \\
        && \\
        \hline
        \hline
        \multirow{4}{*}{\rotatebox{90}{\textcolor{textred}{\makecell[c]{\scriptsize Structure\\-based}}}}& \multirow{2}{*}{\makecell[c]{\textbf{SIT}}} & \multirow{7}{*}{\makecell[l]{Rather than comparing the structure or text of the translation, we leverage the semantics-based \\ measurements to evaluate the \textbf{semantic similarity} (\textbf{or} \textbf{dissimilarity if needed}) of translations, \\ thereby detecting the violation more accurately.}} \\
        && \\
        \cline{2-2}
        &\multirow{2}{*}{\makecell[c]{\textbf{CIT}}}& \\
        && \\
        \cline{1-2}
        \multirow{3}{*}{\rotatebox{90}{\textcolor{textblue}{\makecell[c]{\scriptsize Text\\-based}}}}&\textbf{CAT}& \\
        \cline{2-2}
        &\textbf{Purity}& \\
        \cline{2-2}
        &\textbf{PatInv}& \\
        \hline
    \end{tabular}
\end{table}

Fig.~\ref{overview} presents an overview of our approach. 
The input comprises a pair of source and follow-up input sentences $S_s$ and $S_f$ generated through the input transformation of MRs, as well as their translations $T_s$ and $T_f$ (i.e., source and follow-up outputs) returned by MTS. 
To detect whether MTS violates the MR on the test case pair, 
our approach involves two modules for constructing word closures from the test case pair and one module for detecting the violation by evaluating the semantics of the translation counterparts linked by word closures. 
The construction of word closure requires the word mappings between input sentences and output translations. Thus, in the \textit{first module ``Refined Word Alignment''}, we adopt the word alignment tool and propose two refinement algorithms to build the word alignment between $S_s$ ($S_f$) and $T_s$ ($T_f$). Next, in the \textit{second module ``Word Closure Construction''}, we construct all the word closures from the input test cases to match the finer granularity between the two translations. 
Finally, in the \textit{last module ``Translation Comparison''}, we compare the semantics of the translation fragments matched by the constructed word closures to detect the violations. 
The detected violations are returned as the output. 
In the following, we will illustrate the definition of word closure, the implementation of the three modules, and the finalized metamorphic relation of our approach.

\begin{figure}[t]
\centering
\includegraphics[width=.8\linewidth]{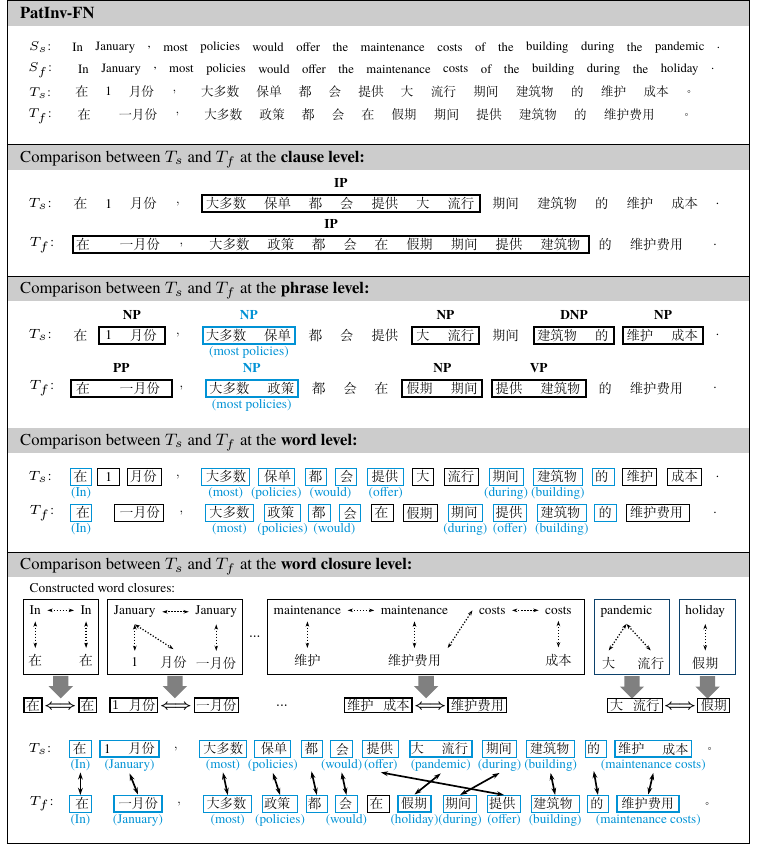}
\caption{Matching between $T_s$ and $T_f$ in PatInv-FN at different granularities. The box represents the granularity of clause, phrase, word, or word closure, in which the blue ones can be matched between $T_s$ and $T_f$. We subdivide the translation into clauses and phrases based on its constituency parse tree generated by Stanford Parser \cite{stanford-parser}, where ``IP'' denotes a clause and ``NP'', ``DNP'', ``PP'', ``VP'' refer to various types of phrases.}
\label{closure-motivation}
\end{figure}

\subsection{Word Closure}
\label{sec:wordClosure}

Before we formally define word closure, we want to show the problems of using native grammar units for comparison. Fig.~\ref{closure-motivation} uses the example of PatInv-FN shown in Fig.~\ref{examples}, where the follow-up input sentence $S_f$ is generated by replacing the word ``pandemic'' in $S_s$ as ``holiday''.
In contrast to PatInv which only requires two outputs to be different, a more reasonable comparison should check whether the \textbf{same input words are translated similarly} and \textbf{the different input words are translated dissimilarly} in $T_s$ and $T_f$. However, to achieve this goal, the first necessary step is to link the counterparts in the source and follow-up outputs.
Now, let us consider the naive method that adopts native grammar units of clauses, phrases, and words represented by the boxes in Fig.~\ref{closure-motivation}.
We identify two problems that motivate us to design a proper granularity for comparison.

The constituency parse tree represents the clause and phrase structures in a sentence. According to the constituency parse tree of $T_s$ and $T_f$, we extract the clauses and phrases of $T_s$ and $T_f$ respectively in Fig.~\ref{closure-motivation}. \textbf{At the clause level}, we can identify one clause in $T_s$ and one clause in $T_f$, both marked as ``IP'' (simple clause). \textbf{At the phrase level}, we can identify five phrases in $T_s$ and four phrases in $T_f$, marked as ``NP'' (noun phrase), ``DNP'' (phrase ended with ``的''), ``PP'' (preposition phrase), and ``VP''(verb phrase). \textbf{At the word level}, a tokenizer is adopted to split $T_s$ and $T_f$ into tokens. 
\textbf{
The first problem is that the grammar units in $\boldsymbol{T_s}$ do not have explicit linkage with their counterparts in $\boldsymbol{T_f}$.}
This lack of linkage creates confusion when determining which units in $T_f$ and $T_f$ respectively should be compared against the MR, particularly for grammar units such as phrases and words that have multiple options. Simply comparing these grammar units between $T_s$ and $T_f$ one by one to detect their difference as the violation is unreasonable as $T_s$ and $T_f$ are not supposed to be totally identical due to different input sentences. 
\textbf{
Another problem is that establishing the links between these grammar units is unpracticable in some cases.} At the clause level, the two clauses in $T_s$ and $T_f$ are not the translation of the same input words and therefore should not be linked for comparison. 
At the phrase level, we can only establish a linkage between the noun phrases (highlighted in blue), i.e., ``大多数 保单'' in $T_s$ and ``大多数 政策'' in $T_f$. For other phrases, their corresponding input words overlap only partially, such as ``1\ 月份'' (January) in $T_s$ and ``在\ 一月份'' (In January) in $T_f$, therefore should not be matched. At the word level, even though more units can match their counterparts in another translation (highlighted in blue), there are still many words that can not be linked as they correspond to different input words.

To address these problems, we propose the concept of ``word closure''. Word closures \textbf{provide a more appropriate granularity of comparison}, meanwhile can \textbf{intrinsically support establishing the necessary linkage} between the fragment in $T_s$ and its counterpart in $T_f$ for output relation checking.
Word closure construction will be performed on the sets of words in $S_s$, $S_f$, $T_s$, and $T_f$, respectively. As a reminder, identical words in the sets of source and follow-up cases should be treated as different objects in our method.

\begin{definition}[Word Closure]
\label{def:closure}
    A word closure $\mathcal{C}$ is an aggregation of four sets of words $[\{w_{i1}^s, w_{i2}^s, \ldots\}, $ $\{t_{j1}^s, t_{j2}^s, \dots\}$, $\{w_{k1}^f, w_{k2}^f, \dots\}, \{t_{t1}^f, t_{t2}^f, \dots\}]$, where $w^s$, $w^f$, $t^s$ and $t^f$ are words from the word sets of the source input $S_s$, follow-up input $S_f$, source translation $T_s$, and follow-up translation $T_f$, respectively.
    Words $w^s$, $w^f$, $t^s$ and $t^f$ can be aggregated in the same word closure, \textit{iff} these words have some correlation with one another, but are not correlated with words outside this closure. As a reminder, it is possible that one or more sets in this aggregation (but not all) are empty, and some words in $T_s$ and $T_f$ are not included in any closure (the detailed construction process will be introduced in Section~\ref{sec:closure}).
    
\end{definition}

By definition, a word closure is a minimal aggregation of linked words, and there is no overlapping among different closures.
For example, the last row of Fig.~\ref{closure-motivation} presents the word closures constructed from PatInv-FN. 
The dotted arrows in the word closures represent the correlations between the words within the input sentences and output translations. 
In the first word closure, the four words, i.e., ``In'' from $S_s$, ``In'' from $S_f$, ``在'' from $T_s$, and ``在'' from $T_s$ have correlations with each other and no correlation with the words outside this closure.
The other four closures also meet these properties, and there is no overlapping among different word closures.
The solid arrows linking $T_s$ and $T_f$ signify the linkages constructed by word closures. It can be observed that word closure establishes the linkage between the fragment in $T_s$ and its counterpart in $T_f$, covering the majority of tokens in both translations. 
Therefore, compared with the native grammar granularities, word closure is more appropriate in identifying violations as the linkages built by word closure enable us to perform a fine-grained output comparison against the output relation.

In the next three sections, we will introduce how to construct word closures for a pair of metamorphic test cases and their outputs, as well as how to compare the outputs against the expected relations. Specifically, we need a refined word alignment to build the correlations between the input and output words, based on which we build closures.

\subsection{Refined Word Alignment}
\label{subsec:wordalignment-refine}
The construction of word closure depends on the word alignment between the input sentences and output translations. 
However, the accuracy of current word alignment tools cannot meet our requirements~\cite{purity, transrepair}: we found that current tools may overlook certain word alignments, resulting in many ineffective word closures that fail to indicate the correct linkage between the source and follow-up output counterparts. 
Thus, we first employ a word alignment tool to initially align the words in the input texts and output translations of the source and follow-up test cases respectively. Then, we match the words in the source and follow-up inputs, and next use two heuristic strategies to refine these missed alignments for better construction of word closures.
Specifically, we first adopt a state-of-the-art word alignment tool, AWESOME, \cite{awesomealign} to build the basic word alignments between $S_s$ ($S_f$) and $T_s$ ($T_f$) and store them in a dictionary $M_s$ ($M_f$). 
For each word $w$ in $S_s$ ($T_s$), AWESOME stores in $M_s[w]$ a set of words $\{w_1, w_2, \ldots\}$ in $T_s$ ($S_s$) that are aligned with $w$. 
$M_f$ functions similarly for $S_f$ and $T_f$.
We then align the unmutated input words in $S_s$ and $S_f$ and store the mapping in the dictionary $M_i$ according to the input transformation. The mutated input words between $S_s$ and $S_f$ are not considered because they are hard to be precisely aligned. 
For each unmutated word $w$ in $S_s$ ($S_f$), $M_i[w]$ returns its corresponding unmutated word $\{w'\}$ in $S_f$ ($S_s$); for each mutate word $w^\prime$, $M_i[w^\prime]$ returns an empty set. 
Next, we apply two strategies to refine the word alignments within $M_s$ and $M_f$, which are built by the word alignment tool.

\begin{figure*}
\centering
\includegraphics[width=0.98\linewidth]{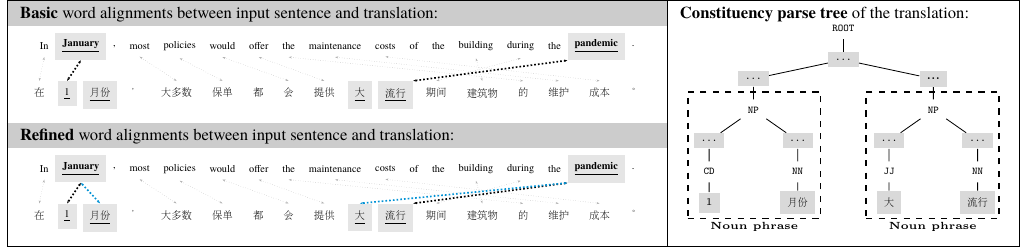}
\caption{Word alignment refinement Algorithm~\ref{alg:one} on PatInv-FN.}
\label{wordalignexp1}
\end{figure*}

\begin{algorithm}[t]
\caption{Alignment Refinement Algorithm 1}
\label{alg:one}
\small
\SetKwInOut{Data}{Data}
\SetKwInOut{Result}{Result}
\Data{$S$: input sentence; $T$: output translation; $Tree$: the constituency parse tree of the translation; $M$: basic word alignments between $S$ and $T$}
\Result{$M$: refined word alignments between $S$ and $T$}
\SetKwFunction{getUncoveredWords}{getUncover}
$T_{unaligned} = \emptyset$\;
\For{each word $w \in T$}
{   
    \If{$M[w]$ is $\emptyset$}{
    $T_{unaligned}\mathrm{.add}(w)$\;
    }
}
\For{each word $w_t \in T_{unaligned}$}
{
    $subtree = \mathrm{getSubtree}(w_t, Tree)$\;
    \If{$subtree$ is a phrase and not a verb phrase}{
        $leaves = \mathrm{getAdjacentLeaves}(subtree, w_t)$\;
        $S_{target} = \mathrm{getAlignedWords}(M, leaves)$\;
        \For{each word $w_s \in S_{target}$}
        {
            $M[w_t]\mathrm{.add}(w_s)$\;
            $M[w_s]\mathrm{.add}(w_t)$\;
        }
    }
}
\Return{$M$}
\end{algorithm}

\textbf{Refinement strategy 1:} 
The first strategy helps to recover the missed word alignments according to the constituency parse tree of the translation. 
If an input word is translated as multiple words in the target language, the multiple output words should be aligned with this input word. However, the word alignment tool may miss some of the alignments between them.
Fig.~\ref{wordalignexp1} shows the alignments unrecognized by the word alignment tool for the example of PatInv-FN. In the example, the ``January'' in the input sentence is translated as two Chinese words ("1" and "月份"), and the input word ``pandemic'' is also translated as two Chinese words (``大'' and ``流行''). However, according to the basic word alignments from AWESOME, ``January'' is only aligned with ``1'' but without ``月份'', and ``pandemic'' is only aligned with ``流行'' but without ``大''. 
In most cases, the multiple translated words corresponding to an input word are always in the same phrase in the translation and function as a whole unit~\cite{phrase}. 
Thus, our refinement strategy 1 is to align the words within the same phrase in the translation to identical input words.
This strategy builds the missing alignments mentioned above, where the phrase structures are recognized based on the constituency parse tree of the translation. As a reminder, we do not consider verb phrases because they may contain too many modifiers to represent very complex semantics~\cite{verbphrase}.

The process of refinement strategy 1 is illustrated in Algorithm~\ref{alg:one}. 
Based on the basic word alignments $M$ between the input sentence $S$ and its translation $T$, we first collect the words in $T$ that are not aligned with any words in $S$ and denote them as $T_{unaligned}$ (Lines 1-6). For each word $w_t$ in $T_{unaligned}$, we locate the smallest subtree $subtree$ that includes both $w_t$ and other words in $T$, based on the constituency parse tree $Tree$ of $T$ (Lines 7-8). If $subtree$ is marked as a phrase and not a verb phrase, we identify the leaf nodes $leaves$ in $subtree$ that are adjacent to $t$ (Lines 9-10). 
Next, we find the input words $S_{target}$ that are aligned with $leaves$, denoted as $S_{target}$ (Line 11). Finally, we build the missed word alignment between $w_t$ and each word $w_s$ in $S_{target}$, which is then added to $M$ (Lines 11-14). 
Take the example shown in Fig.~\ref{wordalignexp1} again. We first collect the words in the translation that are not aligned with any input words as $T_{unaligned}$:\{``月份'', ``都'', ``大''\}.
We then utilize the Stanford parser to obtain the constituency parse tree of the translation, as shown on the right of Fig.~\ref{wordalignexp1}. 
Then, for the first word ``月份'' in $T_{unaligned}$, we locate the smallest subtree marked as a noun phrase comprising the words of ``1'' and ``月份'', as indicated by the dashed box. According to the basic word alignments, ``1'' is aligned with ``January'' but ``月份'' is aligned with nothing. Thus, as indicated by the blue arrow, we build the missed alignment between ``January'' and ``月份''. With similar steps, we also build the missed alignment between ``pandemic'' and ``大''. Finally, following refinement strategy 1, the ``January'' and ``pandemic'' are correctly aligned with their translations ``1'' and ``月份'', and ``大'' and ``流行''.

\begin{figure*}
\centering
\includegraphics[width=\linewidth]{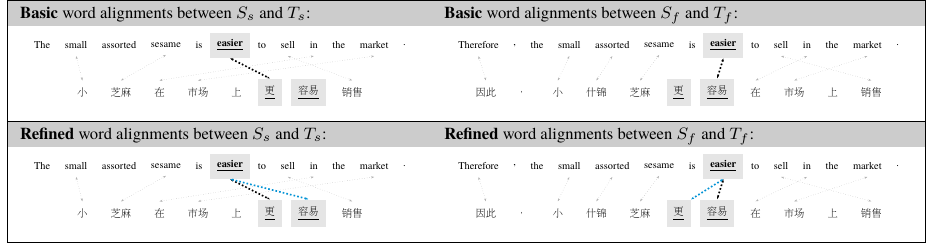}
\caption{Word alignment refinement Algorithm~\ref{alg:two} on CIT-FN.}
\label{wordalignexp2}
\end{figure*}

\textbf{Refinement strategy 2:} 
Our refinement strategy 2 is based on the idea that each unique output word that appears in both $T_s$ and $T_f$ should be aligned with the same input word. 
The same input word is generally translated as the same output words in $T_s$ and $T_f$. 
Therefore, the unique output words shared by $T_s$ and $T_f$ should correspond with the same input word. 
However, the word alignment tool may miss the alignments between the input word and its identical translation in $T_s$ or $T_f$.
As shown in Fig.~\ref{wordalignexp2}, due to the missed word alignment made by AWESOME, the unique output words ``更'' and ``容易'' in $T_s$ and $T_f$ are aligned with different input words. Specifically, the word ``更'' in $T_s$ is aligned with ``easier'' but the word ``更'' in $T_f$ is aligned with nothing. Similarly, the word ``容易'' in $T_f$ is aligned with ``easier'' but the word ``容易'' in $T_s$ is aligned with nothing. Both missed word alignments can not be built by refinement strategy 1 because ``更'' and ``容易'' are included in a verb phrase in both $T_s$ and $T_f$. 
Refinement strategy 2 aims to build the missed alignments of the unique and same output words by comparing the alignments between $S_s$ and $T_s$ and between $S_f$ and $T_f$.

\begin{algorithm}[t]
\caption{Alignment Refinement Algorithm 2}
\label{alg:two}
\small
\SetKwInOut{Data}{Data}
\SetKwInOut{Result}{Result}
\Data{$S_s$: source input sentence; $S_f$: follow-up input sentence; $T_s$: source output translation; $T_f$: follow-up output translation; $M_s$: basic word alignments between $S_s$ and $T_s$; $M_f$: basic word alignments between $S_f$ and $T_f$}
\Result{$M_s$: refined word alignments between $S_s$ and $T_s$; \\$M_f$: refined word alignments between $S_f$ and $T_f$}
$T_{unique} = \mathrm{selectSameAndUniqueToken}(T_s, T_f)$\;
\For{each word $w_t \in T_{unique}$}
{
    \If{$M_s[w_t] != M_f[w_t]$}{
        $S_{target} = M_s[w_t] \cup M_f[w_t]$\;
        \For{each word $w_s \in S_{target}$}
        {
            $M_s[w_t]\mathrm{.add}(w_s)$\;
            $M_s[w_s]\mathrm{.add}(w_t)$\;
            $M_f[w_t]\mathrm{.add}(w_s)$\;
            $M_f[w_s]\mathrm{.add}(w_t)$\;
        }
    }
}
\Return{$M_s$, $M_f$}
\end{algorithm}

Algorithm \ref{alg:two} illustrates the steps of the refinement strategy~2. We first identify all the same and unique words between $T_s$ and $T_f$ as $T_{unique}$ (Line 1). Next, for each word $w_t$ in $T_{unique}$, we compare whether its aligned input words in $S_s$ and $S_f$ are different(Lines 2-3). If its aligned words in $S_s$ and $S_f$ are different, we get the union of them as $S_{target}$ (Line 4). As mentioned above, i.e., the word $w_t$ is supposed to align with the same input words in $S_s$ and $S_f$, we iterate through each word $w_s$ in $S_{target}$ to establish the missed alignment between $w_s$ in $S_s$ ($S_f$) and $w_t$ in $T_s$ ($T_f$), which is then added to $M_s$ ($M_f$) (Lines 5-9).
For instance presented in Fig.~\ref{wordalignexp2}, we initially collect the same and unique words between $T_s$ and $T_f$ as $T_{unique}$:\{``小'', ``芝麻'', ``更'', ``容易'', ``在'', ``市场'', ``上'', ``销售''\}. Then, we traverse each word in $T_{unique}$ to compare its aligned words in $S_s$ and $S_f$. We find that ``更'' is aligned with \{``easier''\} in $S_s$ but is not aligned with any words in $S_f$. Therefore, as marked by the blue arrow, we build the missed alignment between ``easier'' in $S_f$ and ``更'' in $T_f$. With similar steps, we also build the missed alignment between ``容易'' in $T_s$ and ``easier'' in $S_s$. Ultimately, ``easier'' is correctly aligned with its translations ``更'' and ``容易'' in both source and follow-up translations.

\begin{algorithm}
\caption{Word Closure Construction Algorithm}
\label{alg:three}
\small
\SetKwInOut{Data}{Data}
\SetKwInOut{Result}{Result}
\Data{$S_s$: source input sentence; $S_f$: follow-up input sentence; $T_s$: source output translation; $T_f$: follow-up output translation; $M_s$: word alignments between $S_s$ and $T_s$; $M_f$: word alignments between $S_f$ and $T_f$; $M_i$: word alignments between $S_s$ and $S_f$}
\Result{$\mathbb{C}$: all word closures;}
\SetKwFunction{FMain}{closure}
\SetKwFunction{FUnMark}{unmark}
\SetKwFunction{FisMark}{isMark}
\SetKwFunction{FMark}{mark}
\SetKwProg{Fn}{Function}{:}{}
\Fn{\FMain{$C_{sent}^{s}$, $C_{tran}^{s}$, $C_{sent}^{f}$, $C_{tran}^{f}$}}
{
     \While{any change happens to $[C_{sent}^{s}, C_{tran}^{s}, C_{sent}^{f}, C_{tran}^{f}]$}
        {
            $C_{sent}^{f} = \mathrm{getAlignedWords}(M_i, C_{sent}^{s}, S_f)$\;
            $C_{{tran}}^{s} = \mathrm{getAlignedWords}(M_s, C_{sent}^{s}, T_s)$\;
            $C_{{tran}}^{f} = \mathrm{getAlignedWords}(M_f, C_{sent}^{f}, T_f)$\;
            $C_{sent}^{f} = \mathrm{getAlignedWords}(M_f, C_{tran}^{f}, S_f)$\;
            $C_{sent}^{s} = \mathrm{getAlignedWords}(M_i, C_{sent}^{f}, S_s)$\;
            $C_{sent}^{s} = \mathrm{getAlignedWords}(M_s, C_{tran}^{s}, S_s)$\;
        }
    $\mathcal{C} = [C_{sent}^{s}, C_{tran}^{s}, C_{sent}^{f}, C_{tran}^{f}]$\;
    \KwRet{$\mathcal{C}$}\;
}
\textbf{End Function}\\
$\mathbb{C} = \emptyset$\;
\For{each word $w \in S_s \cup S_f$}
    {
        \FUnMark{$w$}\;
    }
\For{each word $w \in S_s$}
{   \If{not \FisMark{$w$}}
    {
        $C_{sent}^{s} = C_{tran}^{s} = C_{sent}^{f} = C_{tran}^{f} = \emptyset$\;
        $C_{sent}^{s}{.add}(w)$\;
        $n = \mathbb{C}.length + 1$\;
        $\mathcal{C}_n =$ \FMain{$C_{sent}^{s}$, $C_{tran}^{s}$, $C_{sent}^{f}$, $C_{tran}^{f}$}\;
        $\mathbb{C}\mathrm{.add}(\mathcal{C}_n)$\;
        \For{each word $w' \in C_{sent}^{s} \cup C_{sent}^{f}$}
        {
            \FMark{$w'$}\;
        }
    }
}
\For{each word $w \in S_f$}
{   \If{not \FisMark{$w$}}
    {
        $C_{sent}^{s} = C_{tran}^{s} = C_{sent}^{f} = C_{tran}^{f} = \emptyset$\;
        $C_{sent}^{f}{.add}(w)$\;
        $n = \mathbb{C}.length + 1$\;
        $\mathcal{C}_n =$ \FMain{$C_{sent}^{s}$, $C_{tran}^{s}$, $C_{sent}^{f}$, $C_{tran}^{f}$}\;
        $\mathbb{C}\mathrm{.add}(\mathcal{C}_n)$\;
        \For{each word $w' \in C_{sent}^{s} \cup C_{sent}^{f}$}
        {
            \FMark{$w'$}\;
        }
    }
}
\Return{$\mathbb{C}$}
\end{algorithm}

\subsection{Word Closure Construction}
\label{sec:closure}

This module constructs all word closures $\mathbb{C}{=}\{\mathcal{C}_1, \mathcal{C}_2, \dots, \mathcal{C}_k\}$ for a metamorphic test case pair. The construction process is illustrated in Algorithm~\ref{alg:three}. 
This process begins with adding one input word from either $S_s$ or $S_f$ to act as the initial word of a new closure. Subsequently, we continually iterate through each word within the closure to add its correlated words within the test cases into the closure until no further new words can be added.
At the very beginning, all words in $S_s$ and $S_f$ are labeled as \textbf{UNMARKED}, which indicates they are not associated with no word closure (Lines 14-15).

We denote the four sets (defined in Definition~\ref{def:closure}) in one single word closure $\mathcal{C}_i$ as $C_{sent}^{s}$, $C_{tran}^{s}$, $C_{sent}^{f}$, and $C_{tran}^{f}$, respectively. Initially, these four sets are empty (Line 19).
The whole construction starts from the first input word in $S_s$ (Line 17). This word is added into $C_{sent}^{s}$ of the very first word closure $\mathcal{C}_1$ (Line 20). Then, we begin to update the four sets by looping through the following six steps within the \texttt{closure} function until no further changes to the four sets (Line 22):
 
\begin{itemize}
    \item[1)] For words in $C_{sent}^{s}$, we assign all their aligned words in $S_f$ to $C_{sent}^{f}$ (Line 3). 
    \item[2)] For words in $C_{sent}^{s}$, we assign all their aligned words in $T_s$ to $C_{tran}^{s}$ (Line 4). 
    \item[3)] For words in $C_{sent}^{f}$, we assign all their aligned words in $T_f$ to $C_{tran}^{f}$ (Line 5). 
    \item[4)] For words in $C_{tran}^{f}$, we assign all their aligned words in $S_f$ to $C_{sent}^{f}$ (Line 6). 
    \item[5)] For words in $C_{sent}^{f}$, we assign all their aligned words in $S_s$ to $C_{sent}^{s}$ (Line 7). 
    \item[6)] For words in $C_{tran}^{s}$, we assign all their aligned words in $S_s$ to $C_{sent}^{s}$ (Line 8). 
\end{itemize}

If there are no further changes to the four sets in the last loop (Line 2), we aggregate these four sets to complete the construction of $\mathcal{C}_1$ (Line 10 and Line 22). After adding $\mathcal{C}_1$ to $\mathbb{C}$ (Lines 23), we label all the words in $C_{sent}^{s}$ and $C_{sent}^{f}$ as \textbf{MARKED} (Lines 24-25). 
Next, we clear these four cache sets and move to the next \textbf{UNMARKED} word in $S_s$ to construct $\mathcal{C}_2$. This process is repeated on each \textbf{UNMARKED} input word in $S_f$ sequentially (Lines 29-39), then, we finish the construction of $\mathbb{C}{=}\{\mathcal{C}_1, \mathcal{C}_2, \dots, \mathcal{C}_k\}$. 
In some cases, certain words in $T_s$ or $T_f$ are not included in any of the word closures due to their lack of correlation with input words. We will illustrate how we deal with the words outside the word closures for output relation comparison in Section~\ref{sec:translation-comparison}.

\begin{figure*}[b]
\centering
\includegraphics[width=1.01\linewidth]{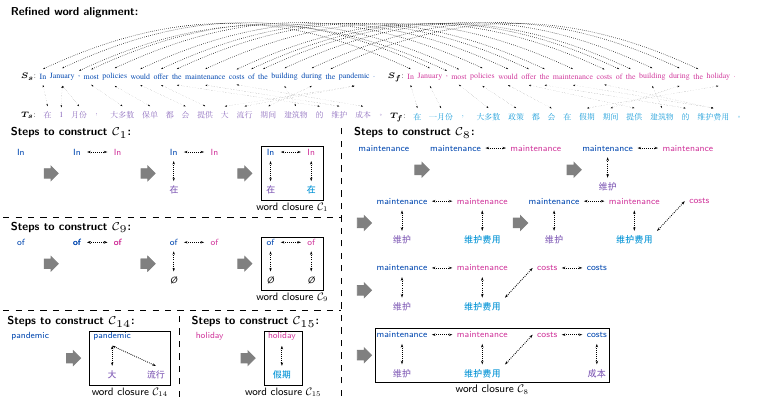}
\caption{Steps to construct word closures from PatInv-FN.}
\label{closureexample}
\end{figure*}
   
The example depicted in Fig.~\ref{closureexample} shows the steps to construct the word closures for the PatInv-FN. 
The constructing process starts from the first word in the source input sentence, i.e., ``In'' in $S_s$. To construct the first word closure $\mathcal{C}_1$, we initialize the four sets in $\mathcal{C}_1$ as $C_{sent}^{s}$:\{In\}, $C_{tran}^{s}{:}\emptyset$, $C_{sent}^{f}{:}\emptyset$ and $C_{tran}^{f}{:}\emptyset$. Then, we update the four sets in $\mathcal{C}_1$ by executing the while loop within \texttt{closure} function. For simplicity, we only show the steps where the four sets have changed:
\begin{itemize}
    \item[1)] For word in $C_{sent}^{s}$:\{In\}, we assign its aligned word in $S_f$, i.e., \{In\}, to $C_{sent}^{f}$. 
    \item[2)] For word in $C_{sent}^{s}$:\{In\}, we assign its aligned word in $T_s$, i.e., \{在\}, to $C_{tran}^{s}$. 
    \item[3)] For word in $C_{sent}^{f}$:\{In\}, we assign its aligned word in $T_f$, i.e., \{在\}, to $C_{tran}^{f}$. 
\end{itemize}
In the last round of updating, no more new words are added to the four sets in $\mathcal{C}_1$. Therefore, the closure $\mathcal{C}_1$ is finalized as [\{In\}, \{在\}, \{In\}, \{在\}] and is added to $\mathbb{C}$. The input words in $\mathcal{C}_1$, namely ``In'' from $S_s$ and ``In'' from $S_f$, are labeled as \textbf{MARKED}. We then move to the next \textbf{UNMARKED} input word to construct the next word closure. The steps to construct $\mathcal{C}_2$, $\mathcal{C}_3$, $\mathcal{C}_4$, $\mathcal{C}_5$, $\mathcal{C}_6$, and $\mathcal{C}_7$ are similar to that for $\mathcal{C}_1$ and thus are omitted for simplicity. After constructing these six closures and labeling the input words within them, namely ``January'', ``most'', ``policies'', ``would'', ``offer'', ``the'' in both $S_s$ and $S_f$, as \textbf{MARKED}, we move to the next \textbf{UNMARKED} input word ``maintenance'' in $S_s$ to construct $\mathcal{C}_8$.

The construction of $\mathcal{C}_8$ is more complex than the previous word closures. We set ``maintenance'' in $S_s$ as the first element and initialize the four sets in $\mathcal{C}_8$ as $C_{sent}^{s}$:\{maintenance\}, $C_{tran}^{s}{:}\emptyset$, $C_{sent}^{f}{:}\emptyset$ and $C_{tran}^{f}{:}\emptyset$. Then, we follow the \texttt{closure} function to update the four sets in $\mathcal{C}_8$. For simplicity, we only present the steps where the four sets have changed:
\begin{itemize}
    \item[1)] For word in $C_{sent}^{s}$:\{maintenance\}, we assign its aligned word in $S_f$, i.e., \{maintenance\}, to $C_{sent}^{f}$. 
    \item[2)] For word in $C_{sent}^{s}$:\{maintenance\}, we assign its aligned word in $T_s$, i.e., \{维护\}, to $C_{tran}^{s}$. 
    \item[3)] For word in $C_{sent}^{f}$:\{maintenance\}, we assign its aligned word in $T_f$, i.e., \{维护费用\}, to $C_{tran}^{f}$. 
    \item[4)] For word in $C_{tran}^{f}$:\{维护费用\}, we assign its aligned words in $S_f$, i.e., \{maintenance, costs\}, to $C_{sent}^{f}$.
    \item[5)] For words in $C_{sent}^{f}$:\{maintenance, costs\}, we assign their aligned words in $S_s$, i.e., \{maintenance, costs\}, to $C_{sent}^{s}$.
    \item[6)] For words in $C_{sent}^{s}$:\{maintenance, costs\}, we assign their aligned words in $T_s$, i.e., \{维护, 成本\}, to $C_{tran}^{f}$.
\end{itemize}
Finally, $\mathcal{C}_8$ is constructed as [\{maintenance, costs\}, \{maintenance, costs\}, \{维护, 成本\}, \{维护费用\}].
Compared with the previous word closures, constructing $\mathcal{C}_8$ requires more updating steps due to more complex word correlations, such as the correlations between ``maintenance'' and ``costs'' in $S_f$ and the word ``维护费用'' in $T_f$. We add $\mathcal{C}_8$ to $\mathbb{C}$ and label the input words in $\mathcal{C}_8$, namely ``maintenance'' and ``costs'' in $S_s$ and ``maintenance'' and ``costs'' in $S_f$, as \textbf{MARKED}.
We then move to the next \textbf{UNMARKED} input word ``of'' in $S_s$ to construct the next word closure $\mathcal{C}_9$. 

$\mathcal{C}_9$ is a special type of word closure, where $C_{tran}^{s}$ or $C_{tran}^{f}$ is empty. We set ``of'' in $S_s$ as the first element and initialize the four sets in $\mathcal{C}_9$ as $C_{sent}^{s}$:\{of\}, $C_{tran}^{s}{:}\emptyset$, $C_{sent}^{f}{:}\emptyset$ and $C_{tran}^{f}{:}\emptyset$. Then, we follow the steps in \texttt{closure} function to update $\mathcal{C}_9$. For simplicity, we only present the steps where the four sets have changed:
\begin{itemize}
    \item[1)] For word in $C_{sent}^{s}$:\{of\}, we assign its aligned word in $S_f$, i.e., \{of\}, to $C_{sent}^{f}$. 
\end{itemize}
As a result, $\mathcal{C}_9$ is constructed as [$C_{sent}^{s}$:\{of\}, $C_{tran}^{s}{:}\emptyset$, $C_{sent}^{f}$:\{of\}, $C_{tran}^{f}{:}\emptyset$]. We add $\mathcal{C}_9$ to $\mathbb{C}$ and label the input words ``of'' in $S_s$ and ``of'' in $S_f$ as \textbf{MARKED}.
We then move to the next \textbf{UNMARKED} input word to construct the next word closure. The steps to build $\mathcal{C}_{10}$, $\mathcal{C}_{11}$, $\mathcal{C}_{12}$, and $\mathcal{C}_{13}$ are similar to previous word closures, thus are omitted for simplicity. After building these four closures and labeling their input words, namely ``the'', ``building'', ``during'', ``the'' in $S_s$ and $S_f$, as \textbf{MARKED}, we move to the next \textbf{UNMARKED} input word ``pandemic'' in $S_s$ to build the closure $\mathcal{C}_{14}$.

Different from the previous input words, the ``pandemic'' is a different input word between $S_s$ and $S_f$. 
We set ``pandemic'' in $S_s$ as the first element and initialize the four sets in $\mathcal{C}_{14}$ as $C_{sent}^{s}$:\{pandemic\}, $C_{tran}^{s}{:}\emptyset$, $C_{sent}^{f}{:}\emptyset$ and $C_{tran}^{f}{:}\emptyset$. Then, we follow the steps in \texttt{closure} function to update $\mathcal{C}_{14}$. For simplicity, we only present the steps where the four sets have changed:
\begin{itemize}
    \item[1)] For word in $C_{sent}^{s}$:\{pandemic\}, we assign its aligned words in $T_s$, i.e., \{大, 流行\}, to $C_{tran}^{s}$. 
\end{itemize}
Finally, $\mathcal{C}_{14}$ is constructed as [$C_{sent}^{s}$:\{pandemic\}, $C_{tran}^{s}$:\{大, 流行\}, $C_{sent}^{f}{:}\emptyset$, $C_{tran}^{f}{:}\emptyset$]. We add $\mathcal{C}_{14}$ to $\mathbb{C}$ and label the input words ``pandemic'' in $S_s$ as \textbf{MARKED}. As all the words in $S_s$ are labeled as \textbf{MARKED}, we move to $S_f$ and find the next \textbf{UNMARKED} input word ``holiday'' in $S_f$ to construct the next word closure $\mathcal{C}_{15}$. The steps to build $\mathcal{C}_{15}$ are similar to that for $\mathcal{C}_{14}$, thus are not presented here for simplicity. Ultimately, all the input words in $S_s$ and $S_f$ are labeled as \textbf{MARKED} and we construct all the word closures from PatInv-FN as $\mathbb{C}{=}\{\mathcal{C}_1, \mathcal{C}_2, \dots, \mathcal{C}_{15}\}$. We will illustrate how to perform the violation checking with these word closures in Section~\ref{sec:translation-comparison}.

\subsection{Translation Comparison}
\label{sec:translation-comparison}

The constructed word closures can be categorized into three types: \textbf{Comparable Word Closure} (CWC), \textbf{Mutated Word Closure} (MWC), and \textbf{Unmatched Word Closure} (UWC). 
Our output relation comparison is performed on these three types of word closures, respectively. 
CWC refers to the word closure that contains the same input words in $S_s$ and $S_f$, as well as their corresponding translations in both $T_s$ and $T_f$. Since the same input words should be translated similarly, we compare whether the translations within each CWC are similar in semantics. 
For detailed examples of CWC, please refer to the ``comparison for CWCs'' illustrated below. 
MWC is the closure that contains the mutated input word in $S_s$ or $S_f$, due to the MR transformation. 
If the mutated words are different in semantics according to the input transformation, we check whether their translations in MWC also have different semantics. Otherwise, we evaluate whether the translations in MWC have similar semantics.
Specific examples of MWC can be found in the following description of the ``comparison for MWCs''.
UWC refers to the word closure that contains the same input words in $S_s$ and $S_f$ but without their translations in $T_s$ or $T_f$.
In other words, UWCs fail to show the linkages between the source and follow-up translations of the same input words. 
However, this does not necessarily indicate that the translations in UWCs are violations because the presence of UWCs could be attributed to the missed word alignments caused by the word alignment tool. If some word alignments are missed, the corresponding translations will not be included in word closures, resulting in UWCs. 
Detailed examples of UWC can be found in the following illustration of `` comparison for UWCs''.
To determine if the same input words in UWCs are translated similarly, we perform a semantic similarity comparison between the translations contained within UWCs and those not included in any word closures.
As a reminder, if the translations in a word closure are all stopwords, 
such as the $\mathcal{C}_{1}$ in Fig.~\ref{closureexample}, where the only translation ``在'' is a stopword, \textbf{we exclude this word closure from the comparison}, considering the negligible information carried by stopwords~\cite{stopwords}.

\textbf{Comparison for CWCs:} 
\label{CWC-comparison}
Each CWC builds the linkage between a fragment in $T_s$ and its counterpart in $T_f$. To check whether the same input words in $S_s$ and $S_f$ are translated similarly in $T_s$ and $T_f$, we perform the semantic similarity comparison for each CWC.
Specifically, within a CWC, we evaluate the semantic similarity score between the words from $T_s$ and the words from $T_f$ and report a violation if their similarity score is below the preset threshold. Take the CWC $\mathcal{C}_{8}$ depicted in Fig.~\ref{closureexample} as example. We compute the semantic similarity between ``维护'', ``成本'' in $T_s$ and ``维护费用'' in $T_f$ and consider them to satisfy the output relation because their similarity score is above the preset threshold.

\textbf{Comparison for MWCs:} 
For MWC, which contains the mutated input words, we perform the semantic dissimilarity (similarity) comparison to check whether the mutated input words, which have different (similar) meanings, are translated dissimilarly (similarly). 
To note, the comparison over MWCs only applies when the transformation ensures that the original and mutated words should be translated similarly or differently. For example, among our five transformations, only IT-5 mutates the original words into another word with a different meaning. As such, we only perform the comparison for MWCs in the test cases generated by IT-5, checking whether the translations in MWCs are dissimilar in semantics.
Among the word closures of PatInv-FN shown in Fig.~\ref{closureexample}, $\mathcal{C}_{14}$ and $\mathcal{C}_{15}$ are MWCs because they contain the mutated input words ``pandemic'' and ``holiday'', which have different meanings. 
For MWCs, we cannot perform comparison within one closure due to the difficulty in aligning the mutated input words, which causes their translations to be scattered across different closures.
Instead, we compare the relations across all the MWCs. Specifically, we collect the translations contained in all the MWCs and evaluate the semantic similarity score between the collected translations from $T_s$ and $T_f$. If the similarity score is above the preset similarity threshold, i.e., violating the requirement for different semantics, we will report a violation. 
So, in Fig.~\ref{closureexample}, we collect the translations from the two MWCs, i.e., ``大'', ``流行'' and ``假期'', for comparison. 
Next, we compute the similarity score between ``大'', ``流行'' from $T_s$ and ``假期'' from $T_f$. Ultimately, we consider them to have different semantics and detect no violation because their similarity score is below the preset threshold.

As a reminder, to support the above comparison practices for CWCs and MWCs, we can use different methods to calculate the similarity between word closures. In this work, we explore five comparison strategies. We suggest the optimal comparison strategy of referring to a synonym database and the language model embeddings. Detailed introduction to these comparison strategies can be found in Section~\ref{sec:rq5}.

\textbf{Comparison for UWCs and all left-over non-stopwords: }
After the comparisons among CWCs and MWCs, we now need to process the comparison for all UWCs, as well as the left-over non-stopwords\footnote{Stopwords are not considered in the comparison.} in $T_s$ and $T_f$ that are not included in any closures. 
Since UWC does not contain the translations from $T_s$ or $T_f$, we can not perform the comparison in each UWC as we do in CWC.
For this part, we use the following operations:

\begin{itemize}
    \item[1)] Collect the non-stopword words in $T_s$ and $T_f$ that are included in UWCs or not included in any word closures and store them to two sets $\mathbb{O}_s$ and $\mathbb{O}_f$, respectively.
    \item[2)] Pair each word in $\mathbb{O}_s$ and each word in $\mathbb{O}_f$ and calculate a total of $|\mathbb{O}_s|\times|\mathbb{O}_f|$ semantic similarity scores.
    \item[3)] Sort these word pairs based on their similarity scores from high to low. 
    \item[4)] Traverse the word pairs sequentially, and match the words that have a similarity score above the threshold and have not been matched before. 
    \item[5)] Report the violation if there are words not matched in $\mathbb{O}_s$ or $\mathbb{O}_f$.
\end{itemize}

\begin{figure*}[h] 
\centering
\includegraphics[width=1.08\linewidth]{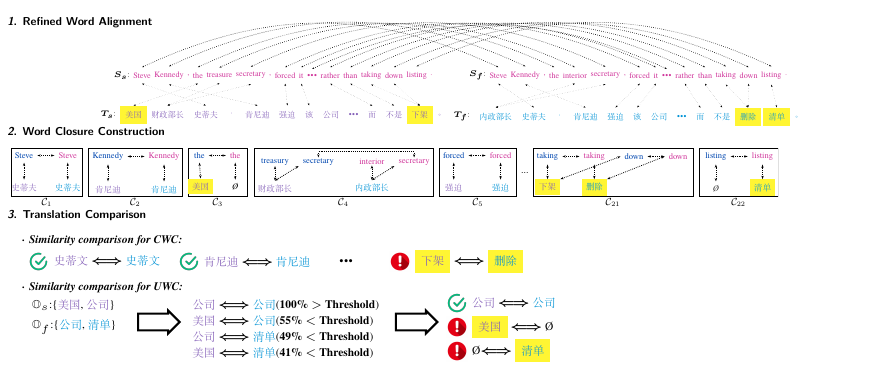}
\caption{The workflow of our approach to resolving the SIT-FN (the translations violate the OR-1 are highlighted in yellow).}
\label{comparisonexp}
\end{figure*}

As all the UWCs of PatInv-FN shown in Fig.~\ref{closureexample} contain no translations or only contain the stopwords, for better illustration, we use the example depicted in Fig.~\ref{comparisonexp} to explain the above comparison steps.
First, we collect the non-stopword translations in the UWCs, i.e. ``美国'' in $\mathcal{C}_3$ and ``清单'' in $\mathcal{C}_{11}$, and the non-stopword translations outside the closures, i.e., ``公司'' in both $T_s$ and $T_f$. Then, we initialize the two sets $\mathbb{O}_s$:\{美国, 公司\} and $\mathbb{O}_f$:\{公司, 清单\}. Next, we calculate the semantic similarity score between each word in $\mathbb{O}_s$:\{美国, 公司\} and each word in $\mathbb{O}_f$:\{公司, 清单\}. After sorting the word pairs according to their similarity score, we match ``公司'' from $T_s$ and ``公司'' from $T_f$ as their similarity score (100\%) is higher than the threshold. The similarity scores between the other word pairs are all below the threshold. As a result, we detect the violation of ``美国'' in $T_s$ and ``清单'' in $T_f$ as they are not semantically matched with the word in $T_f$ and $T_s$ respectively.

In the end, we collect all the violations detected through the above comparison processes for CWCs, MWCs and UWCs and return them as the violation identification results of our method.

\subsection{Metamorphic Relation}

\begin{figure*}[h] 
\centering
\includegraphics[width=0.6\linewidth]{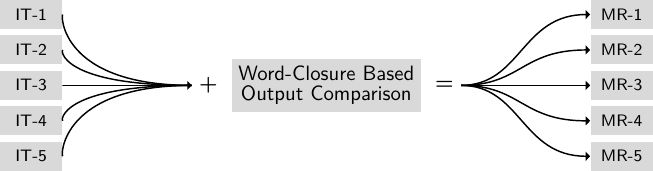}
\caption{Integration of our word-closure based output comparison with existing input transformations.}
\label{newMR}
\end{figure*}

As shown in Fig.~\ref{newMR}, our proposed output relation can be integrated with diverse input transformations to formulate new metamorphic relations. The detailed metamorphic relations are presented in Table~\ref{ourmr}. Given the source and follow-up input sentences generated by any of the five input transformations, our expected output relation is that the same words shared by the two inputs are translated as similar semantics and the changed input words with similar (different) meanings are translated as similar (dissimilar) semantics. 
If the same input words are translated dissimilarly or the changed input words of similar (different) meanings are translated dissimilarly (similarly), at least one of the output translations contains an error and our approach will report a violation.

\begin{table*}[h]
\addtolength{\tabcolsep}{-2.5pt}
\centering\footnotesize
    \caption{Our metamorphic relations for machine translation software.}
    \label{ourmr}
    \begin{tabular}{l|l}
        \hline
        \multicolumn{1}{c|}{\textbf{Input Transformations}} & \multicolumn{1}{c}{\textbf{Output Relation}} \\
        \hline
        \textbf{IT-1:} Replace one word in $S_s$ with another word that keeps the same & 
        \multirow{8}{*}{\makecell[c]{The same words in $\boldsymbol{S_s}$ and $\boldsymbol{S_f}$ should be \\ translated similarly, and the changed words \\ with similar (different) meanings in $\boldsymbol{S_s}$ and $\boldsymbol{S_f}$ \\ should be translated similarly (dissimilarly).}}
        \\
        part of speech to generate $S_f$. & \\
        \cline{1-1}
        \textbf{IT-2:} Replace one word in $S_s$ with another word of similar context- & \\
        aware semantics to generate $S_f$. & \\
        \cline{1-1}
        \textbf{IT-3:} Extract a noun phrase from $S_s$ as $S_f$. &\\
        \cline{1-1}
        \textbf{IT-4:} Insert an adjunct into $S_s$ to generate $S_f$. & \\
        \cline{1-1}
        \textbf{IT-5:} Replace one word in $S_s$ with another word of different meanings, & \\
         or removing a meaningful word or phrase from $S_s$ to generate $S_f$. &\\
        \hline
    \end{tabular}
\end{table*}

\begin{figure*}
\centering
\includegraphics[width=1.0\linewidth]{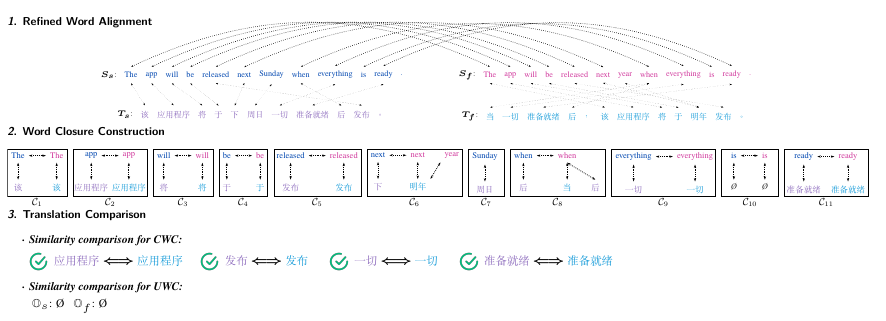}
\caption{The workflow of our approach to resolving the CAT-FP.}
\label{ourmethod-cat}
\end{figure*}

\subsection{Illustration on How Our Approach Addresses Limitations}
\label{sec:examples-solving-limitations}
In this section, we explain how our approach addresses the five limitations of the existing output comparison methods (discussed in Section~\ref{sec:motivation}), with illustrative examples.

\subsubsection{For the limitation of Class-A}\label{sec:examples-solving-limitations-classA}
\textbf{To mitigate this limitation, we propose word closure to achieve finer granularity comparison by building the linkages between the fragment in $T_s$ and its counterpart in $T_f$ for output relation comparison.}
SIT-FN is a false negative of SIT caused by the limitation of Class-A.
Fig.~\ref{comparisonexp} shows how our approach detects the violation in SIT-FN with word closure. 
The input test cases violate the output relation because the same input words ``taking down listing'' in $S_s$ and $S_f$ are translated as different meanings, i.e., ``下架'' (removed from shelves) in $T_s$ and ``删除清单'' (delete list) in $T_f$, and there is a redundant word in $T_s$, i.e., ``美国''(the U.S.).
However, SIT regards this example as non-violation because the overall distance score between the entire $T_s$ and the entire $T_f$ is below the threshold. 
Different from the coarse comparison applied by SIT, our method constructs word closures, $\mathbb{C}{=}\{\mathcal{C}_{1}, \mathcal{C}_{2}, \mathcal{C}_{3}, \dots, \mathcal{C}_{22}\}$, from SIT-FN to build the fine-grained linkages between $T_s$ and $T_f$.
Next, we evaluate the similarity between the fine-grained translation fragments linked by each CWC ($\mathcal{C}_{1}$, $\mathcal{C}_{2}$, $\mathcal{C}_{5}$, ... $\mathcal{C}_{21}$), and successfully detect the violation between ``下架'' and ``删除'' in $\mathcal{C}_{21}$ due to their low similarity at the closure level. Next, by comparing the translations contained in UWCs ($\mathcal{C}_{3}$ and $\mathcal{C}_{22}$) and outside the word closures, we find that the words ``美国'' in $T_s$ and ``清单'' in $T_f$ are not semantically matched with other translations. As a result, after performing the comparison at the word closure level, our approach detects the violation in SIT-FN caused by ``美国'', ``下架'' in $T_s$, and ``删除清单'' in $T_f$.

CAT-FP is a false positive of CAT, also due to the limitation of Class-A.
Fig.~\ref{ourmethod-cat} shows how our approach resolves the CAT-FP with word closure. 
In the example, all the same input words in $S_s$ and $S_f$ are translated similarly in $T_s$ and $T_f$. 
However, since the adverb clauses within $S_s$ and $S_f$, namely ``when everything is ready'', are translated at different positions in $T_s$ and $T_f$, the overall similarity score between $T_s$ and $T_f$ calculated by CAT is very low.
As a result, CAT misidentifies this example as a violation.
Instead of comparing the overall similarity, our method constructs eleven word closures from CAT-FP, $\mathbb{C}{=}\{\mathcal{C}_{1}, \mathcal{C}_{2}, \mathcal{C}_{3}, \dots, \mathcal{C}_{11}\}$, for fine-grained comparison. 
It can be found that the constructed word closure is finer than the adverb clause, which ensures that our translation comparison at the word closure level can avoid being affected by the adverb clause in different positions. After comparing the similarity between the fine-grained fragments linked in each CWC ($\mathcal{C}_{2}$, $\mathcal{C}_{5}$, $\mathcal{C}_{9}$, $\mathcal{C}_{11}$\footnote{$\mathcal{C}_{1}$, $\mathcal{C}_{3}$, $\mathcal{C}_{4}$, $\mathcal{C}_{8}$ in CAT-FP are not compared because the translations that they contain are all stopwords.}), our approach identifies this example as non-violation because all their similarity scores are above the threshold.

\begin{figure*}
\centering
\includegraphics[width=1.0\linewidth]{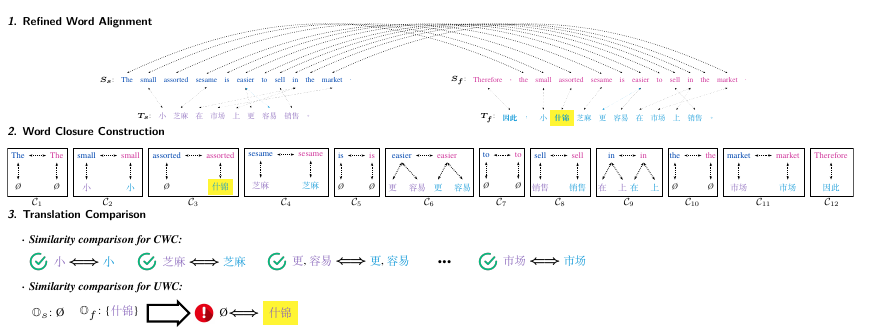}
\caption{The workflow of our approach to resolving the CIT-FN (the translations violate the OR-4 are highlighted in yellow).}
\label{ourmethod-cit}
\end{figure*}

\subsubsection{For the limitation of Class-B}\label{sec:examples-solving-limitations-classB} 
\textbf{This limitation is also addressed by applying word closures, which link the fragment in $\boldsymbol{T_s}$ with its counterpart in $\boldsymbol{T_f}$ and effectively suggest the objects to compare.}
CIT-FN is a false negative of CIT due to the limitation of Class-B. Fig.~\ref{ourmethod-cit} shows how our approach detects the violation in CIT-FN with word closures. In CIT-FN, the same input word ``assorted'' is translated as ``什锦'' (assorted) in $T_s$ but fails to be translated in $T_f$, which violates the output relation. However, due to the lack of links between the words in $T_s$ and $T_f$, CIT applies a loose comparison strategy, i.e., checking whether each path in the constituency parse tree of $T_s$ remains in that of $T_f$, to detect the violation. As a result, CIT can not detect the missing translation in $T_s$ and therefore is unable to detect the violation in this example. 
In our method, we build twelve word closures, $\mathbb{C}{=}\{\mathcal{C}_{1}, \mathcal{C}_{2}, \mathcal{C}_{3}, \dots, \mathcal{C}_{12}\}$, from the test cases. 
It can be found that, in $\mathcal{C}_{3}$, the word ``什锦'' from $T_f$ can not be linked with any word from $T_s$. As a result, our approach detects this violation caused by ``什锦'' because ``什锦'' is not semantically matched with any word from $T_s$ during the comparison for UWC. With the linkages in word closures, we can clearly identify the subjects that should be compared and thus successfully identify this example as a violation.

\begin{figure*}[h]
\centering
\includegraphics[width=1.0\linewidth]{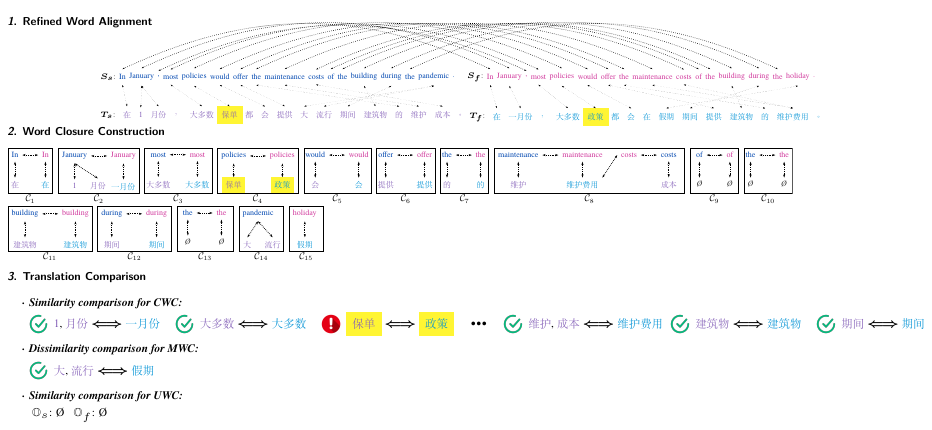}
\caption{The workflow of our approach to resolving the PatInv-FN (the translations violate the OR-5 are highlighted in yellow).}
\label{ourmethod-PatInv}
\end{figure*}

\subsubsection{For the limitation of Class-C}\label{sec:examples-solving-limitations-classC} 
\textbf{We address this limitation by checking not only whether the mutated input words are translated dissimilarly, but also whether the same input words are translated similarly.} 
PatInv-FN is a false negative of PatInv due to the limitation of Class-C. Fig.~\ref{ourmethod-PatInv} illustrates how our approach identifies the violation of PatInv-FN. In the example, the same input word ``policies'' is translated as different meanings in $T_s$ and $T_f$, i.e., ``保单'' (guarantee slip) and ``政策'' (policies). Due to the limitation of Class-C, PatInv is not able to identify the violation caused by the translations of the same input words, therefore failing to detect the violation in this example. 
Our method builds fifteen word closures, $\mathbb{C}{=}\{\mathcal{C}_{1}, \mathcal{C}_{2}, \mathcal{C}_{3}, \dots, \mathcal{C}_{15}\}$, from the test case pair. 
To check whether the same input words are translated similarly, we perform the semantic similarity comparison for each CWC ($\mathcal{C}_{2}$, $\mathcal{C}_{3}$, $\mathcal{C}_{4}$, ..., $\mathcal{C}_{8}$, $\mathcal{C}_{11}$, $\mathcal{C}_{12}$\footnote{The translations in $\mathcal{C}_{1}$ of PatInv-FN are not compared because they are all stopwords.}). As a result, we successfully detect the different semantics between ``保单'' and ``政策'' in $\mathcal{C}_{4}$ and report the violation ignored by PatInv.

\begin{figure*}[h]
\centering
\includegraphics[width=1.07\linewidth]{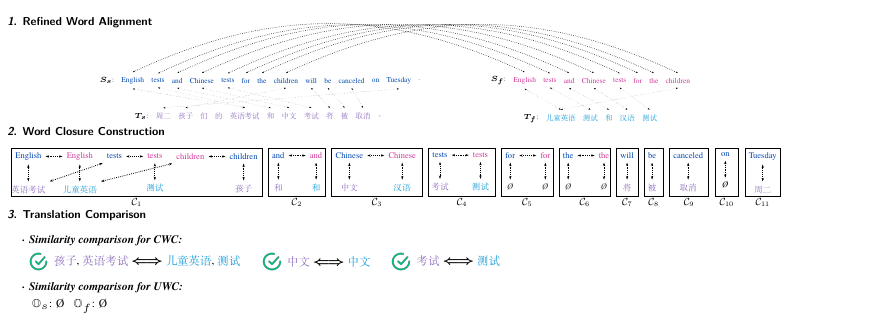}
\caption{The workflow of our approach to resolving the Purity-FP.}
\label{ourmethod-purity}
\end{figure*}

\subsubsection{For the limitations of structure-based and text-based comparison}\label{sec:examples-solving-limitations-textStruc} 
\textbf{Our approach compares the semantics between translations of the matched fragments to address the limitations of both structure-based and text-based comparisons.} 
Semantic comparison is more effective in detecting semantic inconsistencies than structure-based comparison. Additionally, semantic-based measurement can identify synonyms based on their similar meanings, which cannot be recognized by text-based comparison.
Fig.~\ref{ourmethod-purity} shows how our approach resolves the Purity-FP caused by the limitation of text-based comparison. In the example, several input words are translated as different synonyms in $T_s$ and $T_f$, such as the ``Chinese'' is translated as ``中文'' (Chinese) in $T_s$ and ``汉语'' (Chinese) in $T_f$. As a result, unable to recognize the synonyms within the translations, Purity calculates a high distance between $T_s$ and $T_f$ thus misidentifying this case as a violation. Conversely, by measuring the semantic similarity, our approach calculates high similarity scores between the synonyms, 
such as ``孩子\ 英语考试'' and ``儿童英语\ 测试'' in $\mathcal{C}_{1}$, ``中文'' and ``汉语'' in $\mathcal{C}_{3}$, and ``考试'' and ``测试'' in $\mathcal{C}_{4}$\footnote{The translations in $\mathcal{C}_{2}$ of Purity-FP are not compared because they are all stopwords.}. 
Ultimately, our method avoids misidentifying this test case pair as a violation by measuring the semantics.

\section{Experimental Setup}
\label{sec:expsetup}

\subsection{Research Questions}

\textit{\textbf{RQ1: The overall effectiveness of our approach in identifying violations.}} We evaluate the effectiveness of our approach in identifying violations within the metamorphic test case pairs generated by the five input transformations and compare it with the existing output relation comparison methods.

\textit{\textbf{RQ2: Compare with the naive comparison strategies based on native grammar units.}} 
To validate the necessity of adopting word closure, in this RQ, we compare the performance of our approach using word closure and native grammatical units, including clause, phrase, and word.

\textit{\textbf{RQ3: The effectiveness of our approach in locating fine-grained violations.}} The overall violation between a test case pair against the output relation can be attributed to the specific words in the source or follow-up translations, which are denoted as fine-grained violations. Locating the fine-grained violations can speed up the locating and repairing of the errors in MTS. In this RQ, we evaluate the effectiveness of our approach in locating the fine-grained violations. The metric for assessing fine-grained violations is defined in Section~\ref{subsubsec:finegrainmetric}.

\textit{\textbf{RQ4: The efficiency of our approach in identifying violations.}} To understand the practicality of our method, this research question evaluates the efficiency of our approach in terms of time cost and compares it with existing methods.

\textit{\textbf{RQ5: Configuration selection.}} To explore the optimal configuration of our approach, we implement our approach with different configurations, including semantic similarity measurement techniques and threshold for determining the violation. In the end, we conclude the recommended configuration to implement our approach.

\subsection{Systems Under Test}
We select three popular MTSs, namely Google Translate~\cite{google} (\textbf{Google}), Bing Microsoft Translator~\cite{bing} (\textbf{Bing}), and Youdao Translate~\cite{youdao} (\textbf{Youdao}) as the Systems Under Test (SUT).
Specifically, we use the Cloud Translation API from Google Cloud\footnote{\url{https://cloud.google.com/translate/}}, Text Translation API from Microsoft Azure platform\footnote{\url{https://azure.microsoft.com/en-us/products/ai-services/ai-translator/}}, and Text Translation API from Youdao Cloud\footnote{\url{https://ai.youdao.com/product-fanyi-text.s}} to obtain the output translations. 
The translations were collected on February 28, 2023.

\subsection{Methods for Comparison}
We compare our method with the five SOTA MT-based MTS testing methods introduced in Section~\ref{sec:background}, including SIT~\cite{sit}, CAT~\cite{cat}, Purity~\cite{purity}, CIT~\cite{cit}, and PatInv~\cite{patinv}.
We use the codes publicly released by these works to replicate their output comparison methods with the optimal configurations for comparison. It is noteworthy that the performance of SIT, CAT, and Purity can vary with different threshold selections. Therefore, to find the optimal thresholds for the three methods, we execute them across various threshold settings and identify the thresholds that yield the highest F1 scores for comparison.

\subsection{Data Preparation and Annotation}
\label{dataprep}
To evaluate the effectiveness of our method across different languages, we prepare the test cases of two different language pairs, i.e., English-Chinese and Chinese-English. English-Chinese signifies that the input sentence is in English and the output translation is in Chinese. Chinese-English is the inverse. As a reminder, most existing works only study the effectiveness of testing methods in the English-Chinese translation scenario. Since translating text in other languages to English and from Chinese text to other languages is also popular, we study the performance of our method as well as the existing methods on Chinese-English pairs.

To prepare the source input sentence for English-Chinese test cases, we use the dataset constructed by \cite{cit}, which comprises 600 English sentences collected from news articles addressing six topics. As for the source input sentence for Chinese-English test cases, we use News Commentary~\cite{news-commentary} corpus, which is employed by \cite{transrepair} as the source corpus to randomly select 600 source Chinese sentences.

Then we reproduce the five input transformation methods according to their released implementations to generate the follow-up input sentences. Specifically, for each input sentence and each input transformation method, we generate one follow-up sentence to construct a test case pair. There are two implementations available for IT-5, the first one is to replace one word in $S_s$ with a word of different meanings to generate $S_f$, and the other one is to remove a meaningful word or phrase from $S_s$ to generate $S_f$. We opt for the former one to implement IT-5 in our experiment, considering its superior error detection performance compared with another one \cite{patinv}. These input transformation methods inevitably produce sentences with incorrect syntax or that do not meet the input transformation requirements. Therefore, we manually check the generated test cases to remove the unqualified ones. 
Next, we input the test case pairs to get the output translations from the three SUTs. Table~\ref{number} shows the number of the eligible test case pairs for each input transformation and language pair.

\begin{table}[h]
\centering\footnotesize
    \caption{The number of eligible test case pairs for the five input transformations.}
    \label{number}
    \begin{tabular}{ccc}
        \hline
        \textbf{\makecell[c]{Input \\ Transformation}} & \textbf{\makecell[c]{\# English-Chinese\\Test Case Pairs}} & \textbf{\makecell[c]{\# Chinese-English\\Test Case Pairs}} \\
        \hline
        \textbf{IT-1} & 175 & 318 \\
        \hline
        \textbf{IT-2} & 576 & 583 \\
        \hline
        \textbf{IT-3} & 456 & 444 \\
        \hline
        \textbf{IT-4} & 572 & 593 \\
        \hline
        \textbf{IT-5} & 274 & 102 \\
        \hline
    \end{tabular}
\end{table}

To determine if the source and follow-up output translations in each pair of test cases violate the output relation, one of the authors of this paper and a graduate student carried out a manual annotation. Both individuals are proficient in Chinese and English and were provided with comprehensive guidance on the steps to label the violation in advance. They separately labeled all test case pairs, and their results attained a Cohen's Kappa score~\cite{cohen1960coefficient} of 0.83, signifying substantial agreement. Lastly, they deliberated and resolved all the disagreements in their labeling results.

\subsection{Evaluation Metrics}
\subsubsection{Evaluation metrics for violation identification}
To evaluate the effectiveness of our approach in identifying violations, we employ four metrics: \textbf{accuracy}, \textbf{precision}, \textbf{recall}, and \textbf{F1} score. 
Given a pair of test cases violating the MR, a true positive is recorded if our approach correctly identifies it as a violation. Otherwise, a false negative will be recorded.
If our approach identifies a test case pair that does not violate the MR as non-violation, there will be a \textit{true negative}. Conversely, a \textit{false positive} will be accumulated if our approach identifies it as a violation.
Collecting the sum of true positives, true negatives, false positives, and false negatives as $TP$, $TN$, $FP$, and $FN$ respectively, we calculate the four metrics of $accuracy$, $precision$, $recall$, and $F1$ score.

\subsubsection{Evaluation metrics for fine-grained violation locating}
\label{subsubsec:finegrainmetric}
We further evaluate the performance of our approach in locating \textit{the fine-grained violations}, which refer to the specific words in the source and follow-up translations that cause the overall violation of the test case pair. 
Locating fine-grained violations can help developers identify specific translation errors, thereby facilitating the improvement of MTS. Moreover, pinpointing such fine-grained violations can enable users to recognize the suspicious words in the translation provided by MTS during usage.

To assess the effectiveness of our method in locating the fine-grained violations, we define the precision, recall, and F1 score for fine-grained violation locating, which are denoted as $\boldsymbol{Precision_{fine}}$, $\boldsymbol{Recall_{fine}}$ and $\boldsymbol{F1_{fine}}$. Above all, the fine-grained violation $v$ between $T_s$ and $T_f$ can be represented as:
\begin{equation}
    v: [id_{1}^{s}, id_{2}^{s}, ..., id_{1}^{f}, id_{2}^{f}, ...]
\end{equation}
where $id_{x}^{s}$($id_{x}^{f}$) is the index of the word in $T_s$ ($T_f$) that cause the overall violation between $T_s$ and $T_f$.

\begin{figure}[h]
\centering
\includegraphics[width=1\linewidth]{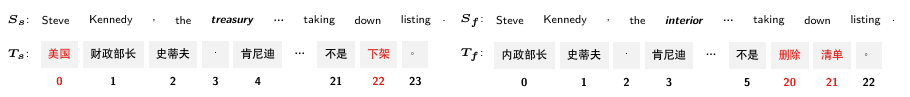}
\caption{The fine-grained violations in SIT-FN.}
\label{fine-grained-example}
\end{figure}

Take the example of SIT-FN depicted in Fig.~\ref{fine-grained-example} to explain how to represent the fine-grained violations. As illustrated in Section~\ref{sec:examples-solving-limitations-classA}, this pair of test cases is a violation due to the four words in $T_s$ and $T_f$, which are marked in red. Therefore, the fine-grained violation $v$ of this example can be recorded as a list of indexes of the four words, i.e., $[0^s, 22^s, 20^f, 21^f]$.

We denote all the test case pairs as $T_{pairs}$. For each test case pair in $T_{pairs}$, we represent the fine-grained violation within the $T_s$ and $T_f$ as $v_x$ and get the fine-grained violation locating result $v_y$ from our approach. 
If an index of $v_y$ exists in $v_x$, a true positive for fine-grained violation locating will be accumulated.
Then, the sum of the true positives on all the test cases, denoted as $TP_{fine}$, is calculated as:

\begin{equation}
    TP_{fine} = \sum_{i=1}^{|T_{pairs}|}match(v_y^i, v_x^i)
\end{equation}
\begin{equation}
    match(v_y, v_x) = \sum_{id_y\in v_y}\sum_{id_x\in v_x} \begin{cases}
        1 & \text{ if } id_y = id_x\\
        0 & \text{ if } id_y \ne id_x
        \end{cases}
\end{equation}

With $TP_{fine}$, the number of false positive, denoted as $FP_{fine}$, and the number of false negative, denoted as $FN_{fine}$, can be derived as follows:

\begin{equation}
    FP_{fine} = \sum_{i=1}^{|T_{pairs}|}|v_y^i| - TP_{fine}
\end{equation}
\begin{equation}
    FN_{fine} = \sum_{i=1}^{|T_{pairs}|}|v_x^i| - TP_{fine}
\end{equation}

Finally, the $Precision_{fine}$, $Recall_{fine}$ and $F1_{fine}$ metrics for evaluating the fine-grained violation locating can be calculated as:
\begin{equation}
    Precision_{fine} = \frac{TP_{fine}}{TP_{fine} + FP_{fine}}
\end{equation}

\begin{equation}
    Recall_{fine} = \frac{TP_{fine}}{TP_{fine} + FN_{fine}}
\end{equation}

\begin{equation}
    F1_{fine} = \frac{2 \cdot Precision_{fine} \cdot Recall_{fine}}{Precision_{fine} + Recall_{fine}}
\end{equation}

\begin{table*}[t]
\centering\footnotesize
    \caption{Evaluation results of the output relation comparison methods on English-Chinese test case pairs.}
    \label{rq1en2zh}
    \begin{tabular}{ccc|rrrrrrrrr}
        \hline
        \multicolumn{1}{c}{\textbf{SUT}} & \textbf{\makecell[c]{Input \\ Transformation}} & \textbf{\makecell[c]{Output \\ Comparison}} & \textbf{\makecell[c]{TP}} & \textbf{\makecell[c]{FP}} & \textbf{\makecell[c]{FN}} & \textbf{\makecell[c]{TN}} & \textbf{\makecell[c]{Acc.}} & \textbf{\makecell[c]{Prec.}} & \ \ \ \textbf{\makecell[c]{Rec.}} & \textbf{\makecell[c]{F1}} & \multicolumn{1}{c}{$\Delta$\textbf{F1}} \\
        \hline
        \multirow{10}{*}{\textbf{Google}} & \multirow{2}{*}{\textbf{IT-1}} & SIT & 16 & 25 & 17 & 117 & 76.0\% & 39.0\% & 48.5\% & 43.2\% & \\
        & & Ours & 28 & 17 & 5 & 125 & \textbf{87.4\%} & \textbf{62.2\%} & \textbf{84.8\%} & \textbf{71.8\%} & \textcolor{textred}{\textbf{+28.6\%}} \\
        \cline{2-12}
        & \multirow{2}{*}{\textbf{IT-2}} & CAT & 47 & 69 & 29 & 431 & 83.0\% & 40.5\% & 61.8\% & 49.0\% & \\
        & & Ours & 60 & 31 & 16 & 469 & \textbf{91.8\%} & \textbf{65.9\%} & \textbf{78.9\%} & \textbf{71.9\%} & \textcolor{textred}{\textbf{+22.9\%}} \\
        \cline{2-12}
        & \multirow{2}{*}{\textbf{IT-3}} & Purity & 60 & 76 & 14 & 306 & 80.3\% & 44.1\% & 81.1\% & 57.1\% & \\
        & & Ours & 61 & 40 & 13 & 342 & \textbf{88.4\%} & \textbf{60.4\%} & \textbf{82.4\%} & \textbf{69.7\%} & \textcolor{textred}{\textbf{+12.6\%}} \\
        \cline{2-12}
        & \multirow{2}{*}{\textbf{IT-4}} & CIT & 58 & 44 & 41 & 429 & 85.1\% & 56.9\% & 58.6\% & 57.7\% & \\
        & & Ours & 80 & 61 & 19 & 412 & \textbf{86.0\%} & 56.7\% & \textbf{80.8\%} & \textbf{66.7\%} & \textcolor{textred}{\textbf{+9.0\%}} \\
        \cline{2-12}
        & \multirow{2}{*}{\textbf{IT-5}} & PatInv & 0 & 0 & 51 & 223 & 81.4\% & 0.0\% & 0.0\% & 0.0\% & \\
        & & Ours & 43 & 29 & 8 & 194 & \textbf{86.5\%} & \textbf{59.7\%} & \textbf{84.3\%} & \textbf{69.9\%} & \textcolor{textred}{\textbf{+69.9\%}} \\
        \hline
        \multirow{10}{*}{\textbf{Bing}} & \multirow{2}{*}{\textbf{IT-1}} & SIT & 7 & 4 & 19 & 145 & 86.9\% & 63.6\% & 26.9\% & 37.8\% & \\
        & & Ours & 21 & 15 & 5 & 134 & \textbf{88.6\%} & 58.3\% & \textbf{80.8\%} & \textbf{67.7\%} & \textcolor{textred}{\textbf{+29.9\%}} \\
        \cline{2-12}
        & \multirow{2}{*}{\textbf{IT-2}} & CAT & 33 & 18 & 39 & 486 & 90.1\% & 64.7\% & 45.8\% & 53.7\% & \\
        & & Ours & 54 & 23 & 18 & 481 & \textbf{92.9\%} & \textbf{70.1\%} & \textbf{75.0\%} & \textbf{72.5\%} & \textcolor{textred}{\textbf{+18.8\%}} \\
        \cline{2-12}
        & \multirow{2}{*}{\textbf{IT-3}} & Purity & 55 & 53 & 11 & 337 & 86.0\% & 50.9\% & 83.3\% & 63.2\% & \\
        & & Ours & 50 & 26 & 16 & 364 & \textbf{90.8\%} & \textbf{65.8\%} & 75.8\% & \textbf{70.4\%} & \textcolor{textred}{\textbf{+7.2\%}} \\
        \cline{2-12}
        & \multirow{2}{*}{\textbf{IT-4}} & CIT & 32 & 31 & 44 & 465 & 86.9\% & 50.8\% & 42.1\% & 46.0\% & \\
        & & Ours & 63 & 37 & 13 & 459 & \textbf{91.3\%} & \textbf{63.0\%} & \textbf{82.9\%} & \textbf{71.6\%} & \textcolor{textred}{\textbf{+25.6\%}} \\
        \cline{2-12}
        & \multirow{2}{*}{\textbf{IT-5}} & PatInv & 0 & 0 & 34 & 240 & 87.6\% & 0.0\% & 0.0\% & 0.0\% & \\
        & & Ours & 28 & 22 & 6 & 218 & \textbf{89.8\%} & \textbf{56.0\%} & \textbf{82.4\%} & \textbf{66.7\%} & \textcolor{textred}{\textbf{+66.7\%}} \\
        \hline
        \multirow{10}{*}{\textbf{Youdao}} & \multirow{2}{*}{\textbf{IT-1}} & SIT & 29 & 48 & 10 & 88 & 66.9\% & 37.7\% & 74.4\% & 50.0\% & \\
        & & Ours & 34 & 23 & 5 & 113 & \textbf{84.0\%} & \textbf{59.6\%} & \textbf{87.2\%} & \textbf{70.8\%} & \textcolor{textred}{\textbf{+20.8\%}} \\
        \cline{2-12}
        & \multirow{2}{*}{\textbf{IT-2}} & CAT & 44 & 61 & 20 & 451 & 85.9\% & 41.9\% & 68.8\% & 52.1\% & \\
        & & Ours & 52 & 38 & 12 & 474 & \textbf{91.3\%} & \textbf{57.8\%} & \textbf{81.2\%} & \textbf{67.5\%} & \textcolor{textred}{\textbf{+15.4\%}} \\
        \cline{2-12}
        & \multirow{2}{*}{\textbf{IT-3}} & Purity & 75 & 80 & 13 & 288 & 79.6\% & 48.4\% & 85.2\% & 61.7\% & \\
        & & Ours & 69 & 42 & 19 & 326 & \textbf{86.6\%} & \textbf{62.2\%} & 78.4\% & \textbf{69.3\%} & \textcolor{textred}{\textbf{+7.6\%}} \\
        \cline{2-12}
        & \multirow{2}{*}{\textbf{IT-4}} & CIT & 56 & 48 & 42 & 426 & 84.3\% & 53.8\% & 57.1\% & 55.0\% & \\
        & & Ours & 81 & 60 & 17 & 414 & \textbf{86.5\%} & \textbf{57.4\%} & \textbf{82.7\%} & \textbf{67.8\%} & \textcolor{textred}{\textbf{+12.8\%}} \\
        \cline{2-12}
        & \multirow{2}{*}{\textbf{IT-5}} & PatInv & 0 & 0 & 53 & 221 & 80.7\% & 0.0\% & 0.0\% & 0.0\% & \\
        & & Ours & 46 & 31 & 7 & 190 & \textbf{86.1\%} & \textbf{59.7\%} & \textbf{86.8\%} & \textbf{70.8\%} & \textcolor{textred}{\textbf{+70.8\%}} \\
        \hline
    \end{tabular}
\end{table*}

\begin{table*}[t]
\centering\footnotesize
    \caption{Evaluation results of the output relation comparison methods on Chinese-English test case pairs.}
    \label{rq1zh2en}
    \begin{tabular}{ccc|rrrrrrrrr}
        \hline
        \multicolumn{1}{c}{\textbf{SUT}} & \textbf{\makecell[c]{Input \\ Transformation}} & \textbf{\makecell[c]{Output \\ Comparison}} & \textbf{\makecell[c]{TP}} & \textbf{\makecell[c]{FP}} & \textbf{\makecell[c]{FN}} & \textbf{\makecell[c]{TN}} & \textbf{\makecell[c]{Acc.}} & \textbf{\makecell[c]{Prec.}} & \ \ \ \textbf{\makecell[c]{Rec.}} & \textbf{\makecell[c]{F1}} & \multicolumn{1}{c}{$\Delta$\textbf{F1}} \\
        \hline
        \multirow{10}{*}{\textbf{Google}} & \multirow{2}{*}{\textbf{IT-1}} & SIT & 37 & 66 & 17 & 198 & 73.9\% & 35.9\% & 68.5\% & 47.1\% & \\
        & & Ours & 43 & 27 & 11 & 237 & \textbf{88.1\%} & \textbf{61.4\%} & \textbf{79.6\%} & \textbf{69.4\%} & \textcolor{textred}{\textbf{+22.3\%}} \\
        \cline{2-12}
        & \multirow{2}{*}{\textbf{IT-2}} & CAT & 65 & 103 & 14 & 401 & 79.9\% & 38.7\% & 82.3\% & 52.6\% & \\
        & & Ours & 48 & 30 & 31 & 474 & \textbf{89.5\%} & \textbf{61.5\%} & 60.8\% & \textbf{61.1\%} & \textcolor{textred}{\textbf{+8.5\%}} \\
        \cline{2-12}
        & \multirow{2}{*}{\textbf{IT-3}} & Purity & 28 & 95 & 18 & 303 & 74.5\% & 22.8\% & 60.9\% & 33.1\% & \\
        & & Ours & 36 & 21 & 10 & 377 & \textbf{93.0\%} & \textbf{63.2\%} & \textbf{78.3\%} & \textbf{69.9\%} & \textcolor{textred}{\textbf{+36.8\%}} \\
        \cline{2-12}
        & \multirow{2}{*}{\textbf{IT-4}} & CIT & 50 & 93 & 49 & 401 & 76.1\% & 35.0\% & 50.5\% & 41.3\% & \\
        & & Ours & 75 & 51 & 24 & 443 & \textbf{87.4\%} & \textbf{59.5\%} & \textbf{75.8\%} & \textbf{66.7\%} & \textcolor{textred}{\textbf{+25.4\%}} \\
        \cline{2-12}
        & \multirow{2}{*}{\textbf{IT-5}} & PatInv & 0 & 0 & 23 & 79 & 77.5\% & 0.0\% & 0.0\% & 0.0\% & \\
        & & Ours & 16 & 8 & 7 & 71 & \textbf{85.3\%} & \textbf{66.7\%} & \textbf{69.6\%} & \textbf{68.1\%} & \textcolor{textred}{\textbf{+68.1\%}} \\
        \hline
        \multirow{10}{*}{\textbf{Bing}} & \multirow{2}{*}{\textbf{IT-1}} & SIT & 35 & 79 & 15 & 189 & 70.4\% & 30.7\% & 70.0\% & 42.7\% & \\
        & & Ours & 36 & 26 & 14 & 242 & \textbf{87.4\%} & \textbf{58.1\%} & \textbf{72.0\%} & \textbf{64.3\%} & \textcolor{textred}{\textbf{+21.6\%}} \\
        \cline{2-12}
        & \multirow{2}{*}{\textbf{IT-2}} & CAT & 45 & 43 & 36 & 459 & 86.4\% & 51.1\% & 55.6\% & 53.3\% & \\
        & & Ours & 52 & 23 & 29 & 479 & \textbf{91.1\%} & \textbf{69.3\%} & \textbf{64.2\%} & \textbf{66.7\%} & \textcolor{textred}{\textbf{+13.4\%}} \\
        \cline{2-12}
        & \multirow{2}{*}{\textbf{IT-3}} & Purity & 35 & 86 & 11 & 312 & 78.2\% & 28.9\% & 76.1\% & 41.9\% & \\
        & & Ours & 35 & 27 & 11 & 371 & \textbf{91.4\%} & \textbf{56.5\%} & 76.1\% & \textbf{64.8\%} & \textcolor{textred}{\textbf{+22.9\%}} \\
        \cline{2-12}
        & \multirow{2}{*}{\textbf{IT-4}} & CIT & 49 & 82 & 51 & 411 & 77.6\% & 37.4\% & 49.0\% & 42.4\% & \\
        & & Ours & 87 & 41 & 13 & 452 & \textbf{90.9\%} & \textbf{68.0\%} & \textbf{87.0\%} & \textbf{76.3\%} & \textcolor{textred}{\textbf{+33.9\%}} \\
        \cline{2-12}
        & \multirow{2}{*}{\textbf{IT-5}} & PatInv & 0 & 0 & 23 & 79 & 77.5\% & 0.0\% & 0.0\% & 0.0\% & \\
        & & Ours & 16 & 8 & 7 & 71 & \textbf{85.3\%} & \textbf{66.7\%} & \textbf{69.6\%} & \textbf{68.1\%} & \textcolor{textred}{\textbf{+68.1\%}} \\
        \hline
        \multirow{10}{*}{\textbf{Youdao}} & \multirow{2}{*}{\textbf{IT-1}} & SIT & 38 & 97 & 11 & 172 & 66.0\% & 28.1\% & 77.6\% & 41.3\% & \\
        & & Ours & 43 & 33 & 6 & 236 & \textbf{87.7\%} & \textbf{56.6\%} & \textbf{87.8\%} & \textbf{68.8\%} & \textcolor{textred}{\textbf{+27.5\%}} \\
        \cline{2-12}
        & \multirow{2}{*}{\textbf{IT-2}} & CAT & 73 & 112 & 18 & 379 & 77.7\% & 39.5\% & 80.2\% & 52.9\% & \\
        & & Ours & 64 & 35 & 27 & 456 & \textbf{89.3\%} & \textbf{64.6\%} & 70.3\% & \textbf{67.4\%} & \textcolor{textred}{\textbf{+14.5\%}} \\
        \cline{2-12}
        & \multirow{2}{*}{\textbf{IT-3}} & Purity & 40 & 38 & 30 & 336 & 84.7\% & 51.3\% & 57.1\% & 54.1\% & \\
        & & Ours & 52 & 20 & 18 & 354 & \textbf{91.4\%} & \textbf{72.2\%} & \textbf{74.3\%} & \textbf{73.2\%} & \textcolor{textred}{\textbf{+19.1\%}} \\
        \cline{2-12}
        & \multirow{2}{*}{\textbf{IT-4}} & CIT & 59 & 93 & 55 & 386 & 75.0\% & 38.8\% & 51.8\% & 44.4\% & \\
        & & Ours & 85 & 56 & 29 & 423 & \textbf{85.7\%} & \textbf{60.3\%} & \textbf{74.6\%} & \textbf{66.7\%} & \textcolor{textred}{\textbf{+22.3\%}} \\
        \cline{2-12}
        & \multirow{2}{*}{\textbf{IT-5}} & PatInv & 0 & 0 & 23 & 79 & 77.5\% & 0.0\% & 0.0\% & 0.0\% & \\
        & & Ours & 19 & 9 & 4 & 70 & \textbf{87.3\%} & \textbf{67.9\%} & \textbf{82.6\%} &\textbf{ 74.5\%} & \textcolor{textred}{\textbf{+74.5\%}} \\
        \hline
    \end{tabular}
\end{table*}

\section{Results and Analysis}
\label{sec:expresult}
\subsection{\textbf{RQ1:} The overall effectiveness of our approach in identifying violations}
\label{sec:rq1}

To evaluate the effectiveness of our approach in identifying violations, we apply our approach to detect the violations from the test cases generated by the five input transformations. After comparing the violation identification results of our approach and the violation labels, we report the four metrics of accuracy, precision, recall, and F1 score, as well as $TP$, $FP$, $FN$, and $TN$. We also report the evaluation results of the output relation comparison methods applied by the five existing works for comparison. 
To ensure impartiality, we present the evaluation results of each existing work with the highest F1 score. Our approach is implemented using the threshold recommended in Section~\ref{sec:rq5}. Table~\ref{rq1en2zh} and Table~\ref{rq1zh2en} respectively present the detailed evaluation results for the English-Chinese and Chinese-English test cases. For clarity, the higher metric scores achieved by our method in comparison to existing methods are highlighted in bold.

Overall, our approach achieves a considerable enhancement in identifying violations compared with the five existing output comparison methods for all the input transformations, SUTs, and language pairs.
For each input transformation, we compare our approach with the existing method specifically designed for this input transformation based on the F1 score. The increments in F1 scores achieved by our approach against the existing methods are presented in the last columns of Tables~\ref{rq1en2zh} and \ref{rq1zh2en} and are highlighted in red. 
As shown in Table~\ref{rq1en2zh}, when detecting violations in the widely investigated English-Chinese translation testing scenario, our approach shows a noteworthy increase of over 10\% in the F1 score in most cases in comparison to the existing methods.
As shown in Table~\ref{rq1zh2en}, in the Chinese-English translation scenario which is not involved by existing works in their experiments and not well resolved by them, our approach also demonstrates much improvement against the five existing methods: it surpasses the existing methods by over 20\% F1 score in most cases.

Next, we analyze how our method improves the performance by addressing the limitations of existing methods:
\begin{itemize}
  \item [1)] 
  As shown in Table~\ref{rq1en2zh} and Table~\ref{rq1zh2en}, the Class-A methods, i.e., SIT and CAT, obtain low precision and recall for IT-1 and IT-2, because of their coarse comparison granularity, which results in many FPs and FNs.
  By contrast, due to the use of word closure as elaborated in Section~\ref{sec:examples-solving-limitations-classA}, our approach achieves higher precision and recall, as marked in bold, and makes fewer FPs and FNs for IT-1 and IT-2 than SIT and CAT in most cases. 
  \item [2)] 
  The Class-B methods, namely Purity and CIT, get low recall for IT-3 and IT-4 due to the lack of linkages between the counterparts within the source and follow-up translations, which leads to many FNs.
  In comparison, our approach makes higher recall, as marked in bold, than both Purity and CIT in most cases with fewer FNs. 
  This is because our approach achieves more rigorous translation comparisons based on the linkages between the translation counterparts built by word closures, as demonstrated in Section~\ref{sec:examples-solving-limitations-classB}.
  \item [3)] 
  For IT-5, the mutated input words are translated dissimilarly in $T_s$ and $T_f$ in all test case pairs. As a result, the Class-C method PatInv \textbf{detects no violations} (i.e. no TPs and no FPs) and gets a zero recall because it only reports violations when the mutated input words are translated similarly. Conversely, our approach also checks whether the unchanged input words are translated similarly in $T_s$ and $T_f$. Therefore, our approach can detect the large amounts of violations missed by PatInv and makes much fewer FNs, resulting in higher recall, as marked in bold.
  \item [4)] 
  As the structure-based comparison methods, SIT and CIT achieve low recall for IT-1 and IT-4, since they are unable to identify the translation counterparts that have different meanings but possess identical syntactic structures, which results in numerous FNs.
  Therefore, the higher recall of our approach for IT-1 and IT-4 can be attributed in part to the application of semantic-based measurement in our approach, which makes fewer FNs as elaborated in Section~\ref{sec:examples-solving-limitations-textStruc}. 
  \item [5)] 
  For IT-2, IT-3 and IT-5, the text-based comparison methods applied by CAT, Purity and PatInv obtain low accuracy or low recall due to the failure to identify synonyms between the source and follow-up translations. 
  Our approach addresses this limitation by assessing the translation semantics to identify the synonyms, achieving higher precision and fewer FPs for IT-2 and IT-3, higher recall and fewer FNs for IT-5.
\end{itemize}

In a word, our approach alleviates the limitations of existing output comparison methods to achieve substantial enhancement in identifying violations.

\subsection{\textbf{RQ2:} Compare with the naive comparison strategies based on native grammar units}

In this RQ, we compare the performance of violation identification based on different granularities, i.e., word closure and three native grammar units (clause, phrase, and word). 

To extract the units of clauses and phrases from the translation, we adopt the constituency parse tree, which represents the clause and phrase structures of a sentence, to guide the extraction process. Specifically, we use the Stanford Parser to parse the source and follow-up translations into their respective constituency parse trees. To extract the clauses, we traverse the constituency parse tree to locate the subtree labeled as a clause and extract its leaves (words) as a clause. Phrase extraction follows the comparable procedure.

\begin{table*}
\centering\footnotesize
    \caption{Violation identification results with different granularities.}
    \label{rq2result}
    \begin{tabular}{ccl|rrrrrrrrl}
        \hline
        \textbf{\makecell[c]{Lang-\\uage}} & \textbf{\makecell[c]{Input \\ Transformation}} & \textbf{Granulariy} & \textbf{TP} & \textbf{FP} & \textbf{FN} & \textbf{TN} & \textbf{Acc.} & \textbf{Prec.} & \ \ \ \textbf{Rec.} & \textbf{\makecell[c]{F1}} & \multicolumn{1}{c}{$\Delta$\textbf{F1}} \\
        \hline
        \multirow{20}{*}{\rotatebox{90}{\textbf{English-Chinese}}} & \multirow{4}{*}{\textbf{IT-1}} & Word Closure& 83 & 55 & 15 & 372 & \textbf{86.7\%} & \textbf{60.1\%} & \textbf{84.7\%} & \textbf{70.3\%} \\
        & & Clause & 26 & 37 & 72 & 390 & 79.2\% & 41.3\% & 26.5\% & 32.3\% & \textcolor{black}{\textbf{-38.0\%}} \\
        & & Phrase & 65 & 142 & 33 & 285 & 66.7\% & 31.4\% & 66.3\% & 42.6\% & \textcolor{black}{\textbf{-27.7\%}} \\
        & & Word & 80 & 113 & 18 & 314 & 75.0\% & 41.5\% & 81.6\% & 55.0\% & \textcolor{black}{\textbf{-15.3\%}} \\
        \cline{2-12}
        & \multirow{4}{*}{\textbf{IT-2}} & Word Closure& 166 & 92 & 46 & 1424 & \textbf{92.0\%} & \textbf{64.3\%} & \textbf{78.3\%} & \textbf{70.6\%} \\
        & & Clause & 70 & 149 & 142 & 1367 & 83.2\% & 32.0\% & 33.0\% & 32.5\% & \textcolor{black}{\textbf{-38.1\%}} \\
        & & Phrase & 110 & 219 & 102 & 1297 & 81.4\% & 33.4\% & 51.9\% & 40.7\% & \textcolor{black}{\textbf{-29.9\%}} \\
        & & Word & 114 & 129 & 98 & 1387 & 86.9\% & 46.9\% & 53.8\% & 50.1\% & \textcolor{black}{\textbf{-20.5\%}} \\
        \cline{2-12}
        & \multirow{4}{*}{\textbf{IT-3}} & Word Closure& 180 & 108 & 48 & 1032 & \textbf{88.6\%} & \textbf{62.5\%} & 78.9\% & \textbf{69.8\%} \\
        & & Clause & 210 & 375 & 18 & 765 & 71.3\% & 35.9\% & \textbf{92.1\%} & 51.7\% & \textcolor{black}{\textbf{-18.1\%}} \\
        & & Phrase & 191 & 193 & 37 & 947 & 83.2\% & 49.7\% & 83.8\% & 62.4\% & \textcolor{black}{\textbf{-7.4\%}} \\
        & & Word & 187 & 233 & 41 & 907 & 80.0\% & 44.5\% & 82.0\% & 57.7\% & \textcolor{black}{\textbf{-12.1\%}} \\
        \cline{2-12}
        & \multirow{4}{*}{\textbf{IT-4}} & Word Closure& 224 & 158 & 49 & 1285 & \textbf{87.9\%} & \textbf{58.6\%} & 82.1\% & \textbf{68.4\%} \\
        & & Clause & 233 & 392 & 40 & 1051 & 74.8\% & 37.3\% & \textbf{85.3\%} & 51.9\% & \textcolor{black}{\textbf{-16.5\%}} \\
        & & Phrase & 221 & 292 & 52 & 1151 & 80.0\% & 43.1\% & 81.0\% & 56.2\% & \textcolor{black}{\textbf{-12.2\%}} \\
        & & Word & 155 & 154 & 118 & 1289 & 84.1\% & 50.2\% & 56.8\% & 53.3\% & \textcolor{black}{\textbf{-15.1\%}} \\
        \cline{2-12}
        & \multirow{4}{*}{\textbf{IT-5}} & Word Closure& 117 & 82 & 21 & 602 & \textbf{87.5\%} & \textbf{58.8\%} & 84.8\% & \textbf{69.4\%} \\
        & & Clause & 137 & 667 & 1 & 17 & 18.7\% & 17.0\% & \textbf{99.3\%} & 29.1\% & \textcolor{black}{\textbf{-40.3\%}} \\
        & & Phrase & 137 & 665 & 1 & 19 & 19.0\% & 17.1\% & \textbf{99.3\%} & 29.1\% & \textcolor{black}{\textbf{-40.3\%}} \\
        & & Word & 88 & 148 & 50 & 536 & 75.9\% & 37.3\% & 63.8\% & 47.1\% & \textcolor{black}{\textbf{-22.3\%}} \\
        \hline
        \hline
        \multirow{20}{*}{\rotatebox{90}{\textbf{Chinese-English}}} & \multirow{4}{*}{\textbf{IT-1}} & Word Closure& 122 & 86 & 31 & 715 & \textbf{87.7\%} & \textbf{58.7\%} & 79.7\% & \textbf{67.6\%} \\
        & & Clause & 35 & 104 & 118 & 697 & 76.7\% & 25.2\% & 22.9\% & 24.0\% & \textcolor{black}{\textbf{-43.6\%}} \\
        & & Phrase & 153 & 801 & 0 & 0 & 16.0\% & 16.0\% & \textbf{100.0\%} & 27.6\% & \textcolor{black}{\textbf{-40.0\%}} \\
        & & Word & 115 & 87 & 38 & 714 & 86.9\% & 56.9\% & 75.2\% & 64.8\% & \textcolor{black}{\textbf{-2.8\%}} \\
        \cline{2-12}
        & \multirow{4}{*}{\textbf{IT-2}} & Word Closure& 164 & 88 & 87 & 1409 & \textbf{90.0\%} & \textbf{65.1\%} & \textbf{65.3\%} & \textbf{65.2\%} \\
        & & Clause & 58 & 103 & 193 & 1394 & 83.1\% & 36.0\% & 23.1\% & 28.2\% & \textcolor{black}{\textbf{-37.0\%}} \\
        & & Phrase & 83 & 177 & 168 & 1320 & 80.3\% & 31.9\% & 33.1\% & 32.5\% & \textcolor{black}{\textbf{-32.7\%}} \\
        & & Word & 159 & 107 & 92 & 1390 & 88.6\% & 59.8\% & 63.3\% & 61.5\% & \textcolor{black}{\textbf{-3.7\%}} \\
        \cline{2-12}
        & \multirow{4}{*}{\textbf{IT-3}} & Word Closure& 123 & 68 & 39 & 1102 & \textbf{92.0\%} & \textbf{64.4\%} & 75.9\% & \textbf{69.7\%} \\
        & & Clause & 144 & 562 & 18 & 608 & 56.5\% & 20.4\% & \textbf{88.9\%} & 33.2\% & \textcolor{black}{\textbf{-36.5\%}} \\
        & & Phrase & 126 & 203 & 36 & 967 & 82.1\% & 38.3\% & 77.8\% & 51.3\% & \textcolor{black}{\textbf{-18.4\%}} \\
        & & Word & 119 & 88 & 43 & 1082 & 90.2\% & 57.5\% & 73.5\% & 64.5\% & \textcolor{black}{\textbf{-5.2\%}} \\
        \cline{2-12}
        & \multirow{4}{*}{\textbf{IT-4}} & Word Closure& 247 & 148 & 66 & 1318 & \textbf{88.0\%} & \textbf{62.5\%} & 78.9\% & \textbf{69.8\%} &  \\
        & & Clause & 261 & 573 & 52 & 893 & 64.9\% & 31.3\% & \textbf{83.4\%} & 45.5\% & \textcolor{black}{\textbf{-24.3\%}} \\
        & & Phrase & 225 & 375 & 88 & 1091 & 74.0\% & 37.5\% & 71.9\% & 49.3\% & \textcolor{black}{\textbf{-20.5\%}} \\
        & & Word & 222 & 164 & 91 & 1302 & 85.7\% & 57.5\% & 70.9\% & 63.5\% & \textcolor{black}{\textbf{-6.3\%}} \\
        \cline{2-12}
        & \multirow{4}{*}{\textbf{IT-5}} & Word Closure& 51 & 25 & 18 & 212 & \textbf{85.9\%} & \textbf{67.1\%} & 73.9\% & \textbf{70.3\%} \\
        & & Clause & 57 & 220 & 12 & 17 & 24.2\% & 20.6\% & 82.6\% & 32.9\% & \textcolor{black}{\textbf{-37.4\%}} \\
        & & Phrase & 68 & 237 & 1 & 0 & 22.2\% & 22.3\% & \textbf{98.6\%} & 36.4\% & \textcolor{black}{\textbf{-33.9\%}} \\
        & & Word & 49 & 29 & 20 & 208 & 84.0\% & 62.8\% & 71.0\% & 66.7\% & \textcolor{black}{\textbf{-3.6\%}} \\
        \hline
    \end{tabular}
\end{table*}

As mentioned in Section~\ref{sec:wordClosure}, we can not simply apply these grammar units since there are no links between the units in $T_s$ and $T_f$, which are demanded for output relation comparison. 
Therefore, we try to build the potential links between these grammar units by adopting the word alignment tool. According to the word alignments, we build the linkage between a unit in $T_s$ and a unit in $T_f$ if the words in the two units are aligned with the same input words. Next, we perform similar translation comparison steps applied in our approach for these grammar units to detect violations. Similar to the comparison for CWC in our approach, we perform the semantic similarity comparison between each pair of linked grammar units. 
For IT-5, to check whether the different input words are translated dissimilarly, we compare the dissimilarity between the units in $T_s$ and the units $T_f$ that are aligned with the different input words. 
For the left units and words in $T_s$ and $T_f$ that are not included in previous comparisons, we apply the same comparison steps for UWC in our approach to detect potential violations within them. 
We merge the test case pairs of the three SUTs together to evaluate the effectiveness of violation identification based on different granularities. 
To make this a fair comparison, we present the evaluation results of the three grammar units with their highest F1 score, and we implement our approach using the threshold recommended in Section~\ref{sec:rq5}.
Table~\ref{rq2result} presents the evaluation results, with the highest metric scores highlighted in bold.

Overall, using word closure surpasses using all the grammar units in violation identification for each input transformation. For English-Chinese test case pairs, word closure achieves significant increases in F1 score compared with three grammar units, particularly on test cases generated by IT-1, IT-2, and IT-5. Among the three grammar units, the unit of word achieves relatively higher F1 scores for IT-1, IT-2, and IT-5, while the phrase unit performs better for IT-3 and IT-4. Given that both IT-3 and IT-4 generate a pair of input sentences with different lengths, it indicates that the phrase unit works better for violation identification between two test cases with different lengths. Meanwhile, our word closure outperforms both phrase and word in terms of F1 scores across all five input transformations. 
As elaborated in Section~\ref{sec:wordClosure}, this is mainly due to the fact that word closures are more flexible than native grammar units to include more words in $T_s$ and $T_f$ for output comparison.

For Chinese-English test cases, word closure still outperforms the three native grammar units in detecting violation, as indicated by the higher F1 scores of word closure, particularly than the clause and phrase units. However, word closure achieves less increase in F1 score than the word unit, within 7\% for all the five input transformations. 
We investigate the reasons for the more outstanding helpfulness of word closure in the English-Chinese test cases. We found that word closure outperforms word unit due to the tokenization ambiguity~\cite{tokenization-ambiguity} between $T_s$ and $T_f$, when the fragment in $T_s$ and its counterpart in $T_f$ that have the same meanings are tokenized into different numbers of words. 
In this case, using the word-level comparison, it is infeasible to establish a one-to-one linkage between every word of the fragment in $T_s$ and every word of its counterpart in $T_f$. Meanwhile, word closures can construct a linkage between the whole fragment in $T_s$ and the counterpart in $T_f$ based on their correlations.
Such tokenization difficulty frequently happens in languages without clear boundaries between words, such as Chinese, Japanese, Thai and so on~\cite{tokenization-difficulty}.
Considering that many low-resource languages also struggle with word tokenization~\cite{low-resource-languages} but can be popularly used as the target translation language (i.e., the language of $T_s$ and $T_f$), it is necessary to adopt word closure in violation identification for these languages.

\subsection{\textbf{RQ3:} The effectiveness of our approach in locating fine-grained violations}

In this RQ, we evaluate the effectiveness of our approach in locating the fine-grained violations.
We compare our approach with the output relation comparison methods used by CAT, Purity, and CIT. We choose these methods because they can pinpoint the precise words in the source or follow-up translations that support their identification of the test case pair as a violation. We pick the test cases of Google Translate for evaluation to bypass repetitive workload that should lead to similar conclusions. For IT-2, IT-3, and IT-4, we collect the true violation test case pairs of Google Translate that have been identified as violations by both our approach and the existing method. Following the same procedure outlined in Section~\ref{dataprep}, we manually annotated the fine-grained violations within these test cases. After comparing the locating results with the annotated labels, we report the $TP_{fine}$, $FP_{fine}$, and $FN_{fine}$, and calculate the three metrics of $Precision_{fine}$, $Recall_{fine}$, and $F1_{fine}$. 
To ensure a fair comparison, we present the evaluation results of CAT, Purity and CIT under their optimal configurations, which are also used for RQ1, and we use the threshold recommended in Section~\ref{sec:rq5} to implement our approach.
Table~\ref{rq3en2zh} and Table~\ref{rq3zh2en} present the evaluation results on English-Chinese and Chinese-English test cases respectively.

\begin{table*}[h]
\centering\footnotesize
    \caption{Fine-grained violation locating result on English-Chinese test case pairs of Google Translate.}
    \label{rq3en2zh}
    \begin{tabular}{cl|rrrccrl}
        \hline
        \textbf{\makecell[c]{Input \\ Transformation}} & \textbf{Approach} & \textbf{$\boldsymbol{\rm{TP_{fine}}}$} & \textbf{$\boldsymbol{\rm{FP_{fine}}}$} & \textbf{$\boldsymbol{\rm{FN_{fine}}}$} & \textbf{$\boldsymbol{\rm{Precision_{fine}}}$} & \textbf{$\boldsymbol{\rm{Recall_{fine}}}$} & \textbf{$\boldsymbol{\rm{F1_{fine}}}$} & \multicolumn{1}{c}{$\Delta\boldsymbol{\rm{F1_{fine}}}$} \\
        \hline
        \multirow{2}{*}{\textbf{IT-2}} & CAT & 76 & 233 & 46 & 24.6\% & 62.3\% & 35.3\% \\
        & Ours & 104 & 19 & 18 & \textbf{84.6\%} & \textbf{85.2\%} & \textbf{84.9\%} & \textcolor{textred}{\textbf{+49.6\%}} \\
        \cline{1-9}
        \multirow{2}{*}{\textbf{IT-3}} & Purity & 88 & 70 & 84 & 55.7\% & 51.2\% & 53.3\% \\
        & Ours & 143 & 20 & 29 & \textbf{87.7\%} & \textbf{83.1\%} & \textbf{85.4\%} & \textcolor{textred}{\textbf{+32.1\%}} \\
        \cline{1-9}
        \multirow{2}{*}{\textbf{IT-4}} & CIT & 49 & 59 & 93 & 45.4\% & 34.5\% & 39.2\% \\
        & Ours & 117 & 30 & 25 & \textbf{79.6\%} & \textbf{82.4\%} & \textbf{81.0\%} & \textcolor{textred}{\textbf{+41.8\%}} \\
        \hline
    \end{tabular}
\end{table*}

\begin{table*}[h]
\centering\footnotesize
    \caption{Fine-grained violation locating result on Chinese-English test case pairs of Google Translate.}
    \label{rq3zh2en}
    \begin{tabular}{cl|rrrccrl}
        \hline
        \textbf{\makecell[c]{Input \\ Transformation}} & \textbf{Approach} & \textbf{$\boldsymbol{\rm{TP_{fine}}}$} & \textbf{$\boldsymbol{\rm{FP_{fine}}}$} & \textbf{$\boldsymbol{\rm{FN_{fine}}}$} & \textbf{$\boldsymbol{\rm{Precision_{fine}}}$} & \textbf{$\boldsymbol{\rm{Recall_{fine}}}$} & \textbf{$\boldsymbol{\rm{F1_{fine}}}$} & \multicolumn{1}{c}{$\Delta\boldsymbol{\rm{F1_{fine}}}$} \\
        \hline
        \multirow{2}{*}{\textbf{IT-2}} & CAT & 85 & 354 & 41 & 19.4\% & 67.5\% & 30.1\% \\
        & Ours & 80 & 27 & 46 & \textbf{74.8\%} & 63.5\% & \textbf{68.7\%} & \textcolor{textred}{\textbf{+38.6\%}} \\
        \cline{1-9}
        \multirow{2}{*}{\textbf{IT-3}} & Purity & 53 & 62 & 16 & 46.1\% & 76.8\% & 57.6\% \\
        & Ours & 49 & 7 & 20 & \textbf{87.5\%} & 71.0\% & \textbf{78.4\%} & \textcolor{textred}{\textbf{+20.8\%}} \\
        \cline{1-9}
        \multirow{2}{*}{\textbf{IT-4}} & CIT & 44 & 41 & 64 & 51.8\% & 40.7\% & 45.6\% \\
        & Ours & 79 & 15 & 29 & \textbf{84.0\%} & \textbf{73.1\%} & \textbf{78.2\%} & \textcolor{textred}{\textbf{+32.6\%}} \\
        \hline
    \end{tabular}
\end{table*}

Overall, our approach significantly outperforms the three existing methods in locating fine-grained violations. This indicates that the violation identification result of our approach is based on more accurate fine-grained violation locating results, and therefore is more dependable. Specifically, while detecting violations in English-Chinese test cases, our approach demonstrates a superior $F1_{fine}$ score of more than 30\% compared with other methods. As for the Chinese-English test cases, our approach maintains a 30\% improvement in the $F1_{fine}$ score for both CAT and CIT while surpassing Purity by 20.8\%.

Next, we provide an in-depth analysis of how our approach improves upon the three existing methods:
\begin{itemize}
  \item [1)] Our approach outperformed CAT by significantly reducing $FP_{fine}$. Specifically, we achieve a reduction in $FP_{fine}$ from 233 to 19 for English-Chinese test cases, and a reduction from 354 to 27 for Chinese-English test cases. This improvement is attributable to the finer granularity we applied for output relation comparison, which is determined by word closure. Given the pairs of fine-grained fragments linked by word closures, it is possible to identify which one violates the output relation, which is unattainable for CAT.
  \item [2)] Our approach gets considerably lower $FP_{fine}$ when compared with Purity. Specifically, our approach makes 20 and 7 false negatives for English-Chinese and Chinese-English respectively, whereas Purity has 70 and 62 false negatives. This improvement is mainly attributed to the fact that Purity fails to recognize the synonyms within translations, thereby misidentifying them as violations. Conversely, our approach utilizes semantic measurement to decrease such false negatives because synonyms exhibit high semantic similarity. 
  \item [3)] For IT-4, the $TP_{fine}$ of our approach is almost double the $TP_{fine}$ of CIT. The lower $TP_{fine}$ of CIT can be attributed to two limitations. The first one is that CIT applies a loose comparison method due to the lack of linkages between the fragments in $T_s$ and $T_f$. More specifically, CIT examines if each word in $T_s$ satisfies the output relation OR-4, as detailed in Table~\ref{taxonomy}. If any word in $T_s$ is found to violate the OR-4, CIT reports it as a fine-grained violation. As a result, CIT is limited to locating the fine-grained violations in $T_s$ rather than in $T_f$, while our approach can locate the fine-grained violations in both $T_s$ and $T_f$ due to the rigorous comparison we realize. The second limitation is that, as a structure-based comparison method, CIT cannot identify the violations with identical syntactic structure but different meanings. Our approach evaluates semantics to detect violations to eliminate this limitation. Therefore, our approach achieves a significantly higher $TP_{fine}$ than CIT.
\end{itemize}

\subsection{\textbf{RQ4:} The efficiency of our approach in identifying violations}
\label{sec:rq4}
In this RQ, we study the efficiency of our approach in identifying violations and compare it with the five existing methods. For test cases generated by each input transformation, we run each method individually five times using the same experimental settings and present the average time cost per test case pair as the final result. Table~\ref{rq4efficiency} presents the average running time of our approach and the five existing methods per test case pair for five input transformations on Google Translate. We use the evaluation results on Google Translate as representative because the results on other SUTs exhibit similar trends.

\begin{table*}[h]
\centering\footnotesize
    \caption{Efficiency of the output relation comparison methods on test case pairs of Google Translate.}
    \label{rq4efficiency}
    \begin{tabular}{cc|c|c|c|c}
        \hline
        \multirow{2}{*}{\textbf{\makecell[c]{Input \\ Transformation}}} & \multirow{2}{*}{\textbf{\makecell[c]{Output \\ Comparison}}}  & \multicolumn{2}{c|}{\textbf{English-Chinese}} & \multicolumn{2}{c}{\textbf{Chinese-English}} \\
        \cline{3-6}
        & & \textbf{Average Time Cost} & $\Delta$ & \textbf{Average Time Cost} & $\Delta$\\
        \hline
        \multirow{2}{*}{\textbf{IT-1}} & SIT & 0.066s & & 0.487s &\\
         & Ours & 0.436s & +0.370s & 0.809s & +0.322s\\
        \hline
        \multirow{2}{*}{\textbf{IT-2}} & CAT & 0.054s & & 0.168s &\\
         & Ours & 0.444s  & +0.390s & 0.806s & +0.638s\\
        \hline
        \multirow{2}{*}{\textbf{IT-3}} & Purity & 0.002s & & 0.190s &\\
         & Ours & 0.190s & +0.188s & 0.358s & +0.168s\\
        \hline
        \multirow{2}{*}{\textbf{IT-4}} & CIT & 0.399s & & 0.179s &\\
         & Ours & 0.380s & -0.019s & 0.747s & +0.568s\\
        \hline
        \multirow{2}{*}{\textbf{IT-5}} & PatInv & 0.002s & & 0.007s &\\
         & Ours & 0.446s & +0.444s & 0.677s & +0.670s\\
        \hline
    \end{tabular}
\end{table*}

As shown in Table~\ref{rq4efficiency}, the average running time of our approach per test case pair is relatively higher than that of the existing methods. However, the absolute time cost of our method is actually very low (ranging from 0.190s to 0.809s).
Besides, we can find that the extra time cost of our method ranges from 0.168s to 0.670s, \textbf{which is also very low and practically can be negligible as compared with general human perceptual speed}.

Actually, the additional time cost is mainly attributed to the application of multiple NLP techniques, such as the word alignment tool (which is employed in our approach to enhance the identification of translation errors) and semantic similarity evaluation (which is employed in our approach to enhance the measurement of text semantics), etc.
Despite the additional time required, our approach achieves significantly higher effectiveness than the baseline methods in identifying translation errors, which will benefit the subsequent process of fixing (discussed in Section~\ref{subsec:fixing}). So \textbf{it is reasonable to believe that such additional time overhead is worthwhile.}

\subsection{\textbf{RQ5:} Configuration selection}
\label{sec:rq5}

In this RQ, we evaluate the performance of our approach under various configurations. 
The previous evaluations of our approach are all performed with the recommended optimal configuration in this RQ.
Specifically, our approach allows for two configuration options: the semantic similarity measurement and the corresponding threshold for determining violations.
Initially, we implement our approach with three commonly used semantic similarity measurements~\cite{semantic-similarity}:

\begin{itemize}
\item{\textbf{Database for synonyms} (\textbf{Config-1}):} The database for synonyms consists of an extensive collection of words with their synonyms. 
In our method, if two sets of words being compared are recognized as synonyms to one another according to the synonym database, we consider them to have similar semantics. If not, they are deemed dissimilar in semantics. 
Specifically, given two sets of words $\mathcal{W}_i=\{w_1^i, w_2^i, \ldots, w_m^i\}$ and $\mathcal{W}_j=\{w_1^j, w_2^j, \ldots, w_n^j\}$ for comparison, we first store all the synonyms retrieved from the synonym database for each word in $\mathcal{W}_i$ into the set $\mathcal{S}_i=\{s_1^i, s_2^i, \ldots, s_x^i\}$ and store all the synonyms of each word in $\mathcal{W}_j$ into the set $\mathcal{S}_j=\{s_1^j, s_2^j, \ldots, s_y^j\}$ from the synonym database. Next, we consider $\mathcal{W}_i$ and $\mathcal{W}_j$ having similar semantics only if $\mathcal{S}_i$ is a subset of $\mathcal{S}_j$ or $\mathcal{S}_j$ is a subset of $\mathcal{S}_i$.
In our experiments, we use the famous synonym database WordNet~\cite{wordnet} to retrieve synonyms for both Chinese and English words. 

\item{\textbf{Cosine similarity of word vectors} (\textbf{Config-2}):} Word vectors, which are learned from large corpora, can capture the semantic relationship between words~\cite{word-vector}. 
Our method identifies two pieces of translations as semantically similar if the cosine similarity score between their vector representations exceeds the preset threshold, while those with a similarity score below the threshold are deemed dissimilar. 
Specifically, given two sets of words $\mathcal{W}_i=\{w_1^i, w_2^i, \ldots, w_m^i\}$ and $\mathcal{W}_j=\{w_1^j, w_2^j, \ldots, w_n^j\}$ for comparison, we collect the word vectors for each word in $\mathcal{W}_i$ ($\mathcal{W}_j$) and calculate their average vector $V_i$ ($V_j$), which is a common practice in NLP. Next, we consider $\mathcal{W}_i$ and $\mathcal{W}_j$ to have similar semantics if the cosine similarity between $V_i$ and $V_j$ is higher than the preset threshold.
In our experiments, we employ the Word2Vec model~\cite{word2vec} trained on Google News corpus to obtain English word vectors and use the Chinese word vectors that have been trained by \cite{Chinese-word-vector} on 9 large Chinese corpora.

\item{\textbf{Cosine similarity with large language model} (\textbf{Config-3}):} 
The semantics of a word can vary depending on its sentence context~\cite{semantics-context}. Compared with the static word vectors used in Config-2, the word embeddings generated by the large language model take into account the specific context in which the word appears, therefore capturing the word's semantics more precisely~\cite{contextual-word-embeddings,FCS-bert-context-information}.
Similar to Config-2, translations are considered semantically similar if their embeddings exhibit a cosine similarity score above the threshold. Otherwise, we consider them dissimilar. Specifically, we adopt the English version and the Chinese version of the Bert-base model~\cite{bert} to extract word vectors for English and Chinese words respectively. 
\end{itemize}

In addition, we set another two configurations by combining the above approaches:

\begin{itemize}
\item{\textbf{Config-4}}: 
We utilize a combination of synonym database and cosine similarity of word vectors to evaluate the semantic similarity. If either of the two measurements indicates that the translations being compared are similar in semantics, we identify them as semantically similar. Otherwise, we consider the translations to have dissimilar semantics.

\item{\textbf{Config-5}}: We employ a combination of synonym database and cosine similarity with large language model to evaluate the semantic similarity. Similar to Config-4, we only identify translations being compared as dissimilar in semantics if both measurements suggest so. Otherwise, we consider them as similar in semantics.
\end{itemize}

\begin{figure}[t]
\centering
\includegraphics[width=.7\linewidth]{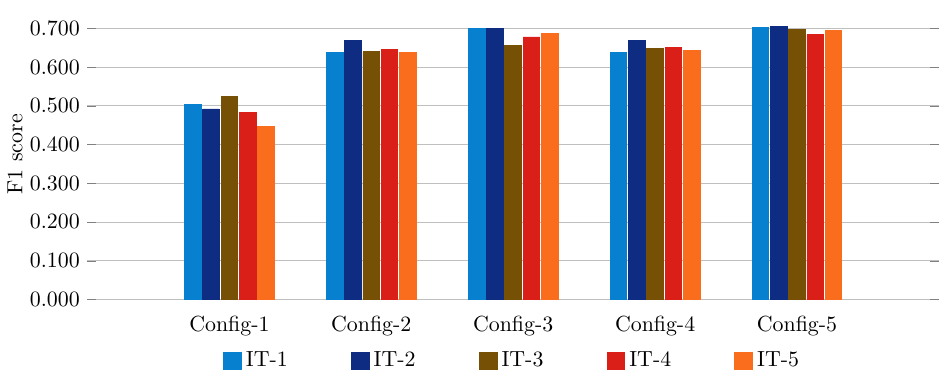}
\caption{Effectiveness with different semantic measurements for English-Chinese test case pairs.}
\label{confien2zh}
\end{figure}

\begin{figure}[t]
\centering
\includegraphics[width=.7\linewidth]{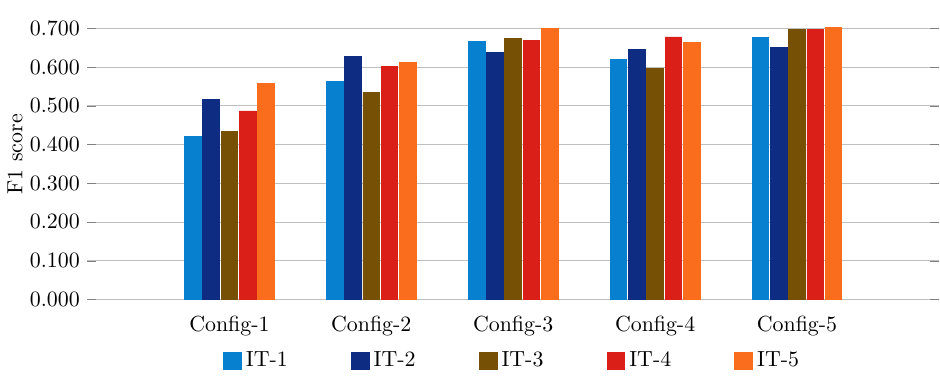}
\caption{Effectiveness with different semantic measurements for Chinese-English test case pairs.}
\label{configzh2en}
\end{figure}

We implement our approach employing the above five configurations and evaluate its performance on all the test cases generated by each input transformation. The best F1 scores achieved by our approach on the English-Chinese and Chinese-English test cases are shown in Fig.~\ref{confien2zh} and Fig.~\ref{configzh2en}, respectively. Notably, the five configurations display a similar distribution of F1 scores for both language pairs.

Among Config-1, Config-2, and Config-3, which use only one measurement approach, Config-3 achieves relatively higher F1 scores across all five input transformations. 
This suggests the effectiveness of the word vectors generated by the large language model in evaluating semantic similarity. Config-4 and Config-5, which combine two semantic measurements, achieve slightly higher F1 scores than Config-2 and Config-3 respectively. 
It indicates that the combination of the synonym database and vector-based cosine similarity can achieve better performance than utilizing them individually. 
On the whole, we advocate implementing our approach with Config-5, which employs both the synonym database and cosine similarity with large language model for evaluating semantic similarity.

\begin{figure}[t]
\centering
\includegraphics[width=.7\linewidth]{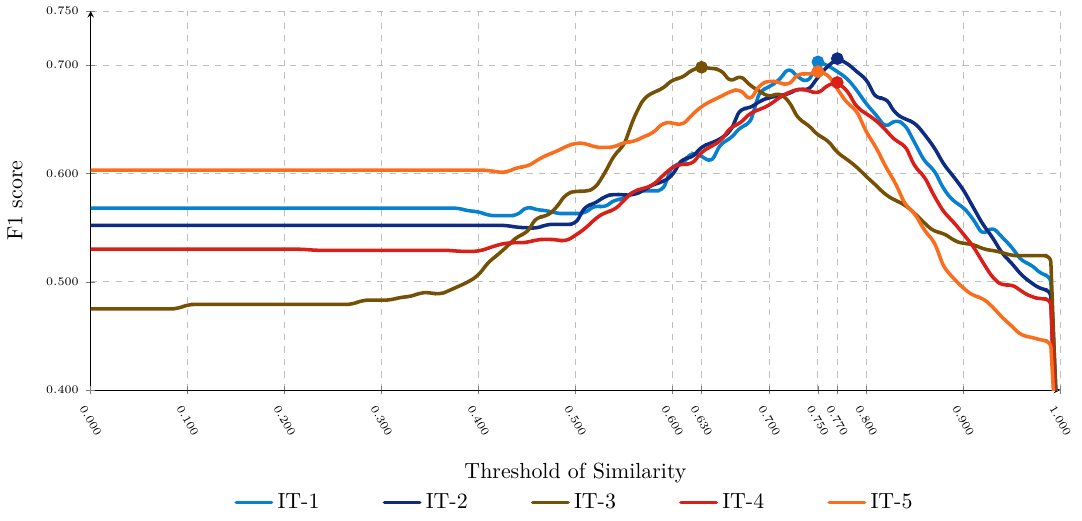}
\caption{Effectiveness over different similarity thresholds for English-Chinese test case pairs.}
\label{thresholden2zh}
\end{figure}

\begin{figure}[t]
\centering
\includegraphics[width=.7\linewidth]{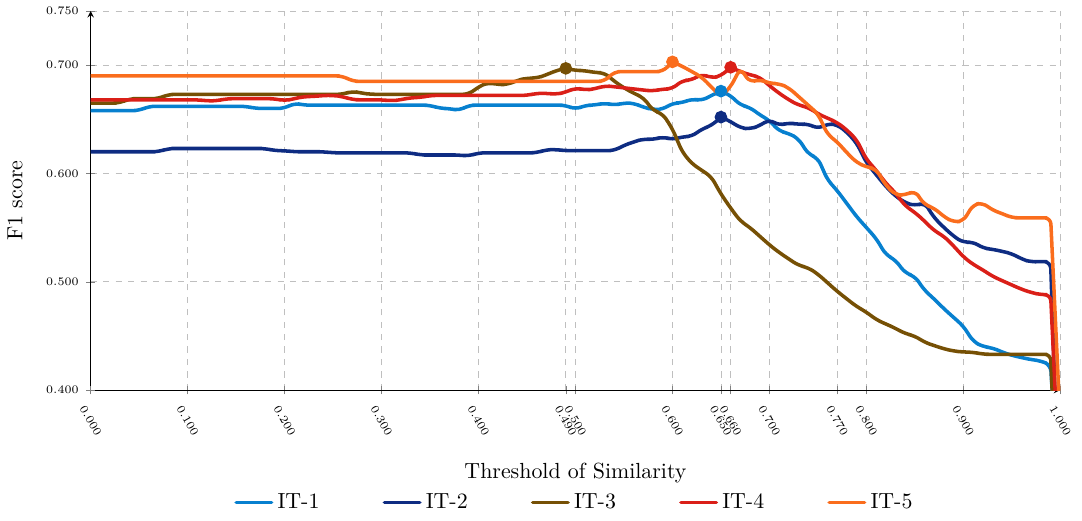}
\caption{Effectiveness over different similarity thresholds for Chinese-English test case pairs.}
\label{thresholdzh2en}
\end{figure}

For Config-5, we proceed to evaluate our approach with different thresholds and conclude the threshold that demonstrates the best performance. We set the threshold from 0 to 1 in the step of 0.01 and record the F1 score of our approach. Fig.~\ref{thresholden2zh} and Fig.~\ref{thresholdzh2en} present the F1 scores of our approach with different thresholds on English-Chinese and Chinese-English test cases respectively.
According to the highest F1-scores presented in Fig.~\ref{thresholden2zh}, the suggested thresholds for IT-1, IT-2, IT-3, IT-4, and IT-5 are 0.75, 0.77, 0.63, 0.77, and 0.75, respectively, while detecting violations in English-Chinese test cases.
While identifying violations in Chinese-English test cases, as shown in Fig.~\ref{thresholdzh2en}, the suggested thresholds for IT-1, IT-2, IT-3, IT-4, and IT-5 are 0.65, 0.65, 0.49, 0.66, and 0.60 respectively. 

\section{Discussion}
\label{sec:discussion}

In this section, we offer an analysis beyond the scope of our research questions to show the helpfulness of our approach in improving the robustness of MTS models. Additionally, we discuss the limitations of our method and list potential directions for future improvements.

\subsection{Helpfulness of Our Approach in Improving the Quality of MTS}
\label{subsec:fixing}

The above experimental evaluation has shown the advantages of our method in revealing translation errors. In this section, we would like to discuss how helpful our method is in improving the robustness of MTS models. 
In this discussion, we simulate a typical semi-automatic repairing practice with fine-tuning technique, by strictly following the process in the existing works of He et al.~\cite{sit,purity} and Gupta et al.~\cite{patinv}. In this repairing practice, they pick up a sample translation model, and run a small set of test cases with their method to report violations. The output pairs that violate their comparison method are then passed to human participants to further label the true translation errors. These annotated true errors will be used in fine-tuning, expecting that they can be properly fixed.

By following this process, we pick up mBART~\cite{mBART}, as the subject MTS model, and randomly sample 200 test input pairs (constructed with transformations IT-1 to IT-5) from our dataset for testing, with 100 pairs for English-Chinese and another 100 pairs for Chinese-English (each transformation has equal number of input pairs). 
mBART is pre-trained on large-scale monolingual corpora in many languages and achieves state-of-the-art in translating between many language pairs. We use the English and Chinese sentence pairs from a famous multilingual corpus, OPUS-100~\cite{opus}, to fine-tune the mBART model\footnote{We apply the checkpoint of ``facebook/mbart-large-50'' released on the Hugging Face model hub for fine-tuning.} to further improve its performance in translating English to Chinese and Chinese to English. 
Next, we run the 200 test input pairs on mBART, and check the output relations with our method and the baseline methods, respectively.
The testing results are then passed to participants who are familiar with both Chinese and English to manually label the true erroneous translations among the revealed violations.
Finally, we fine-tune the machine translation model with the labeled data\footnote{After fine-tuning with the labeled data for two epochs, the BLEU scores of mBART on the OPUS-100 test set remain relatively stable, with declines lower than 0.0065.}.

\begin{table*}[h]
\centering\footnotesize
    \caption{Comparison between our method and the baselines in improving the quality of MTS.}
    \label{finetune-timecost}
    \begin{tabular}{c|c|c|c|c|c|c}
        \hline
        \textbf{\makecell[c]{Language}} & \multicolumn{3}{c|}{\textbf{English-Chinese}} & \multicolumn{3}{c}{\textbf{Chinese-English}} \\
        \hline
        \textbf{\makecell[c]{Testing Approach}} & \makecell[c]{Five MRs with\\Baselines} & \makecell[c]{Five MRs with\\Our Method} & $\Delta$ & \makecell[c]{Five MRs with\\Baselines} & \makecell[c]{Five MRs with\\Our Method} & $\Delta$ \\
        \hline
        \textbf{\makecell[c]{\# Identified\\Translation Errors}} & 19 & 24 & +5 & 18 & 20 & +2 \\
        \hline
        \textbf{\makecell[c]{Successful Rate\\of Repairing}} & 89.47\% & 95.83\% & +6.36\%  & 94.44\% & 95.00\% & +0.56\% \\
        \hline
        \textbf{\makecell[c]{Time Cost for\\Annotating Data}} & 786s & 434s & -352s & 838s & 431s & -407s \\
        \hline
    \end{tabular}
\end{table*}

Table~\ref{finetune-timecost} presents the repairing results based on the five metamorphic relations integrated with our word-closure based comparison method and with the baseline comparison methods, respectively.
Row 3 shows the number of identified translation errors after manual annotation; while Row 4 shows the successful rate of repairing, i.e., the proportion of successfully fixed translation errors in the identified ones. 
It can be found that the confirmed translation errors (Row 3) identified among violations reported by our comparison method are more than those by the baselines. Besides, our method achieves relatively higher successful rates (Row 4) of repairing than the baselines.

Actually, by inspecting the process of this semi-automatic repairing practice, it can be found that the quality of manual annotations is very important. It is true that we can recruit reliable volunteers, but the quality of the raw testing results passed to these volunteers will affect the difficulty of labeling activity. Evidence to show the degree of difficulty is the time cost of the labeling process. 
We recorded the time cost of this step with our method and with the baselines, which are shown in Row 5 of Table~\ref{finetune-timecost}. It can be found that identifying true errors among the detected violations by our method is \textbf{much faster} than that by the baselines (the time saving is 352s and 407s for English-Chinese and Chinese-English, respectively). 
By communicating with the participants, they gave us an intuitive explanation that can support the above argument: Our method can significantly reduce the false alarms among the revealed violations and locate the errors at a finer granularity, with which the manual annotation can be much more efficient and convenient. 
In other words, the better effectiveness of our method in identifying violations can benefit the fixing process as well.

\subsection{Limitations and Future Improvement Directions of Our Approach}

In this section, we first summarize the main limitations of our approach to explain the inaccurate cases made by our approach and then propose several possible directions that may help improve the effectiveness and efficiency of our approach.

The current limitation of our approach mainly comes from the false alarms and missed alarms in the testing results. To deeply inspect the reasons, we conduct a manual analysis. Following the manual analysis, we found that most defects of our approach can be attributed to the intrinsic limitations of the NLP techniques that it builds upon, such as the word alignment tool and the semantic measurement techniques. Regarding the word alignment tool, although we have proposed several heuristic strategies to refine some missed word alignments, there is still a potential risk that this tool may lead to inaccurate linkages between translations, which may further result in improper word closures. Similarly, for the semantic comparison method, even though we have experimented with diverse semantic measurement techniques and selected the optimal one, it may still produce erroneous assessments in some complex situations. 
For instance, given a source translation ``In the eyes of many in the region ...'' and its follow-up translation ``... has been seen by many in the region ...'', the two texts ``In the eyes of'' and ``been seen by'', which share identical meanings, should be compared according to our output relation. However, the semantic comparison model adopted in our method assigns them a low similarity score of 0.64, thus incorrectly identifying them as different semantics, leading to a false alarm.
For another instance, given two Chinese words, ``貂皮'' and ``水貂'', which mean ``fur of mink'' and ``mink'' respectively and should be compared according to our output relation, the semantic comparison model adopted in our method assigns them a high similarity score of 0.80, thus incorrectly identifying them as the same semantics, leading to a missed alarm.

To further enhance the effectiveness and efficiency of our current approach, we suggest several potential directions that may help to eliminate the limitations of our approach for future advancements. In terms of effectiveness, since our approach is built upon several existing NLP techniques, incorporating a more enhanced word alignment tool and semantic measurement techniques in the future may prove an effective means of enhancing the effectiveness of our method. When it comes to efficiency, given that our approach comprises multiple separable processing steps, it would be beneficial to improve efficiency by executing some processing steps concurrently, thereby decreasing the overall time cost.

\section{Threats to Validity}
\label{sec:threatstovalidity}

The first threat to validity lies in the manual assessment of the violation identification results. To mitigate this threat, one of the authors of this paper and a graduate student manually inspected each pair of test cases individually and achieved a Cohen's Kappa score of 0.83. They are all proficient in both Chinese and English. Ultimately, they discussed the inconsistencies between their inspections and reached a uniform result.

The second threat to validity comes from the word alignment tool employed in our approach. The precision of the word alignment results may affect the construction of word closures and further impact the performance of our approach in identifying violations. 
To minimize the inaccuracies caused by the word alignments, we adopt the state-of-the-art word alignment tool, namely AWESOME, to build the word alignments. In addition, we propose two heuristic strategies to refine the word alignment results, which further improves their correctness and completeness.

The last threat to validity is about the generalizability of our approach. To alleviate this threat, on one hand, we evaluate our approach on three popular and representative MTSs (Google Translate, Bing Microsoft Translator and Youdao Translate). 
On the other hand, we did not only evaluate our method on the English-to-Chinese translation task as the baseline works~\cite{sit,purity,cit,cat} did, but also evaluated it on the Chinese-to-English translation task. 
Actually, our method can be easily adapted to diverse languages. 
Given a new translation task from Languages $L_A$ to $L_B$, we can still use the same process and current word alignment tool (which is based on a multilingual model that supports 104 languages) in Section~\ref{subsec:wordalignment-refine} to link the related tokens. 
The only adaptation is to replace current text-parsing tools and the database for synonyms with the ones for new languages, which shall be convenient and straightforward. 
Once the relations are built, the following procedure to construct word closures and perform output comparison will have no difference from the ones shown in Sections~\ref{sec:closure} and \ref{sec:translation-comparison}.

\section{Related Work}
\label{sec:relatedwork}
\subsection{Machine Translation Testing}
Traditional testing for MTS is performed with parallel language corpora.
The quality of MTS is evaluated by comparing its output with the reference translation in the corpora. 
Although many automatic evaluation metrics, such as Blue~\cite{blue} and Rouge~\cite{rouge}, have been proposed to conduct the comparison automatically, obtaining high-quality parallel evaluation datasets is still a challenging issue~\cite{parallel-corpus}.

Many testing methods that do not rely on reference translation have been proposed for MTS. Initially, MTS is tested via round-trip translation~\cite{RTT2005,RTT2006,RTT2007}. To improve the similarity comparison for round-trip translation, Cao et al.~\cite{semmt} propose the SemMT testing method, which assesses the semantics similarity between two translations based on their regular expressions. Pesu et al.~\cite{MT-Monte-Carlo} test MTS using a Monte Carlo method that detects incorrect translations by comparing the consistency of translation results generated by different translating paths. Wang et al.~\cite{phrase-mapping} develop a testing method to detect the under- and over-translation according to the bilingual mappings learned from parallel language datasets. After that, many metamorphic testing methods that involve input transformation were proposed. Sun et al.~\cite{MT4MT} introduced a method that replaces a word in a short sentence and detects translation errors by checking whether more than one word differ in its translation. He et al. propose two metamorphic testing methods, namely SIT~\cite{sit} and Purity~\cite{purity}. SIT identifies erroneous translations by replacing one word within a sentence and measuring the structure variance in its translation. Purity checks whether the noun phrases are translated similarly in different contexts to detect translation errors. Gupta et al.~\cite{patinv} propose the testing method named PatInv, which deletes or replaces one word in a sentence and checks whether the sentence's translation has changed. Sun et al. propose TransRepair~\cite{transrepair} and CAT~\cite{cat} to test MTS via word replacement. Both TransRepair and CAT detect incorrect translations by replacing a single word within the input sentence and checking whether the other unchanged input words are translated consistently. The distinction between the two methods is that CAT improves the word replacement of TransRepair by making the replacement context-aware. Ji et al.~\cite{cit} propose the testing method CIT to test MTS with constituency invariance relation. CIT detects translation errors by checking whether the constituency parse tree of a sentence's translation is retained after inserting an adjunct into the sentence. 
In this paper, to alleviate the limitations of the five latest metamorphic testing methods for MTS, which involve input transformations, we propose the word closure-based output comparison method to achieve more accurate and finer violation identification.

\subsection{Metamorphic Testing for Natural Language Processing}
With the fast development of Natural Language Processing (NLP), the testing for NLP software has received much attention. 
However, due to the large input space for complex NLP tasks and the absence of existing labels~\cite{ML-testing}, one of the challenges faced in testing NLP software is the Oracle Problem~\cite{oracle-problem}.
Among the software testing approaches, metamorphic testing is found to be an effective approach for addressing the oracle problem~\cite{MT_Review}.
In addition to machine translation, some existing works leverage metamorphic testing to test various applications in natural language processing, including sentiment analysis~\cite{sentiment-testing2,checklist,sentiment-testing,sentiment-testing3}, natural language understanding~\cite{NLI-testing,NLU-testing}, named entity recognition~\cite{NER-testing}, part of speech tagging~\cite{POS-testing}, textual content moderation~\cite{CM-testing}, machine reading comprehension~\cite{checklist,MRC-testing} and question answering system~\cite{checklist,QA-tesing1,QA-tesing3,QA-tesing4,QA-tesing2}. These works normally propose novel MRs with new input transformation methods to test the SUT. In this paper, we aim to improve the output relation comparison method of existing works.

\section{Conclusion and Future Work}
\label{sec:conclusion}
In this paper, we propose a metamorphic testing approach for MTSs based on the word closure. Specifically, we propose a new comparison unit, i.e., word closure, which indicates the linkages between the fine-grained counterparts in the source and follow-up translations. During testing, we first build word closures for each group of source and follow-up inputs and outputs, and next measure the semantic similarity (or dissimilarity if needed) between each pair of linked fragments to identify the incorrect translations. Based on word closure, our approach compares the fine-grained counterparts between the source and follow-up output translations against the output relation to address the limitations of existing works in mistranslation detection and achieve more effective violation detection.
The evaluation shows that our approach is more effective than the five latest existing methods in translation violation detection, achieving an average increase of 29.9\% in the F1 score. Besides, our approach outperforms existing works with an average increase of 35.9\% in F1 score in locating specific fine-grained violations. 
In the future, we plan to expand our word closure-based output comparison method to more diverse input transformations. 
We will also investigate the usefulness of the fine-grained violation locations in repairing translation errors.

\section*{Acknowledgments}
This work was supported by National Natural Science Foundation of China (Grant No. 62250610224). We sincerely appreciate the valuable suggestions from the anonymous reviewers for our paper.

\clearpage
\end{CJK*}
\bibliographystyle{ACM-Reference-Format}
\bibliography{ref}

\end{document}